\pdfoutput=1
\documentclass[twocolumn,prx,aps,showpacs]{revtex4-1}
\usepackage{amsmath}
\usepackage{amssymb}
\usepackage{graphicx}
\usepackage{esint}
\usepackage[colorlinks=true,citecolor=blue, linkcolor=blue]{hyperref}

\graphicspath{{./Fig/}}

\begin{document}

\title{The New Phase due to Symmetry Protected Piecewise Berry Phases; Enhanced Pumping and Non-reciprocity in Trimer Lattices}

\author{Xuele Liu}
\altaffiliation{xuele@okstate.edu}

\affiliation{120 W Miller Ave, Stillwater, Oklahoma 74078, USA}

\author{G.S. Agarwal}
\altaffiliation{On leave from The Department of Physics, Oklahoma State University, Stillwater, Oklahoma 74078, USA}

\affiliation{Institute for Quantum Science and Engineering, Department of Biological
and Agricultural Engineering, Texas A\&M University, College Station,
TX 77845}

\date{\today}
\begin{abstract}
Finding new phase is a fundamental task in physics. Landau\textquoteright s
theory explained the deep connection between symmetry breaking and
phase transition commonly occurring in magnetic, superconducting and
superfluid systems. The discovery of the quantum Hall effect led to
$\mathbb{Z}$ topological phases which could be different for same
symmetry and are characterized by the discrete values of the Berry
phases. By studying 1D trimer lattices we report new phases characterized
by Berry phases which are piecewise continuous rather than discrete
numbers. The phase transition occurs at the discontinuity point. With
time-dependent changes, trimer lattices also give a 2D phases characterized
by very specific 2D Berry phases of half period. These Berry phases
change smoothly within a phase while change discontinuously at the
transition point. We further demonstrate the existence of adiabatic
pumping for each phase and gain assisted enhanced pumping. The non-reciprocity of the pumping process makes the system a good optical diode.
\end{abstract}
\maketitle

\section{Introduction}

The discovery of the Berry phase \cite{Berry1980,Berry1984,WilczekZee1984}
and the theory of Quantum Hall effect \cite{TKNN1982,Kohmoto1984},
has led to large number of studies on the topological states of matter.
Three distinct properties characterize non-interacting topological
states of matter. These are the Berry phase, discrete symmetry and
band gap between the energy bands in parameter space. In translationally
invariant systems, Bloch momentum $\mathbf{k}$ is the parameter and
the Brillouin zone is the parameter space. The Berry phase is then
the external phase acquired by the eigenstate $\psi_{n}\left(\mathbf{k}\right)$
of Hamiltonian $\mathcal{H}\left(\mathbf{k}\right)$ while parameter
$\mathbf{k}$ changes adiabatically around a loop in the Brillouin
zone. For electrons, the physics is determined by the filled energy
bands. A \textit{characteristic parameter} can be defined based on
the Berry phases of the filled bands \cite{TKNN1982,Kohmoto1984,Class,Class1,Class2,Alexandra2014,Alexandra2016}.
The discrete symmetry of system allows only discrete values of the
characteristic parameter \cite{Class,Class1,Class2,Alexandra2014}.
Each discrete value relates to a specific topological structure (a
complete ball, a complete torus, e.t.c.) and is called the topological
number \cite{HasanReview2010}. Continuous changes of parameters of
the Hamiltonian may continuously deform the energy bands, however
it can not change the characteristic parameter, unless the band gap
is closed and reopened to form a new type of band structure \cite{HasanReview2010,QiReview2011}.
In this sense, different matter phases are labeled by the discrete
topological numbers. These topological phases have distinct physical
properties such as type and number of robust edge modes and the corresponding
quantum electric transport \cite{shenbook2012,bernevigbook2013,XL2012}.
Any perturbation with respect to the symmetry which preserves the
band gap can not destroy the phase \cite{Hua2009}. The symmetry is
important as when it is broken, the system can be smoothly changed
from one phase to another without closing the gap \cite{nagaosa2013}.

For the quasi-1D system, each filled energy band $\varepsilon_{i}\left(k\right)$
has the Berry phases (module $2\pi$) $\theta_{i}\in\left[-\pi,\pi\right]$.
The quantity $\frac{\theta_{i}}{2\pi}=\left[-\frac{1}{2},\frac{1}{2}\right]$
gives the position of the Wannier center of the corresponding (hybrid)
energy band, i.e. the center of mass of electrons in each unit cell,
here the size of unit cell is supposed to be $1$ \cite{Rui2011,Vanderbilt2014,Alexandra2014,Vanderbilt2015}.
Obviously, non-zero Berry phase $\theta_{i}\neq0$ means the center
of mass of electrons is not same as the center of mass of atoms, the
system is polarized. For 1D sub-lattice system \cite{Class,Class1,Class2},
sum of Berry phases of filled band is a characteristic parameter.
Non-trivial topology of such a system only allows $\theta_{i}=\pm\pi$,
which means maximum non-zero polarization of the system. When the
the system is finite i.e. has boundaries, the polarization is reflected
by the occurrence of extra edge eigenstates, for which electrons are
localized at the boundary \cite{bernevigbook2013}.

The 2D nontrivial topology leads to an important new aspect which
is called adiabatic pumping \cite{Thouless1983} by the way of dimensional
reduction \cite{Qi2008}. To understand this, let us fix $k_{y}$
in $\mathcal{H}\left(k_{x},k_{y}\right)$ and get the polarization
$\frac{\theta_{i}\left(k_{y}\right)}{2\pi}$ along $x$-direction,
here $\theta_{i}\left(k_{y}\right)$ is a function of $k_{y}$. Then
we adiabatically change $k_{y}$ by one period from $-\pi\sim\pi$
and measure the changes of polarization $\frac{\theta_{i}\left(k_{y}\right)}{2\pi}$.
Non-trivial 2D ($\mathbb{Z}$) topology means the center of mass of
electrons $\frac{\theta_{i}\left(k_{y}\right)}{2\pi}$ adiabatically
changes from $\frac{1}{2}$ (the most right of the unit cell) to $-\frac{1}{2}$
(the most left), and vice versa. When the system is finite along $x$-direction
and has boundary, this process is equivalent to the edge band slowly
merging into bulk band and then reappearing in the gap \cite{Qi2008}.
Correspondingly, the eigenstate slowly changes from localized state
at one edge to the bulk state and then to the localized state at another
edge. An integer number of electrons from all the filled bands are
adiabatically pumped from one edge to another during the period. It
should be mentioned, for one fixed $k_{y}$ , $\mathcal{H}\left(k_{x},k_{y}\right)$
in general is no longer a 1D topology insulator, or else $\frac{\theta_{i}\left(k_{y}\right)}{2\pi}$
is quantized and can not smoothly change while $k_{y}$ changes.

\begin{figure*}[t]
\begin{minipage}[t]{0.95\textwidth}%
\begin{minipage}[b]{0.55\textwidth}%
\begin{flushleft}
\includegraphics[width=0.96\textwidth]{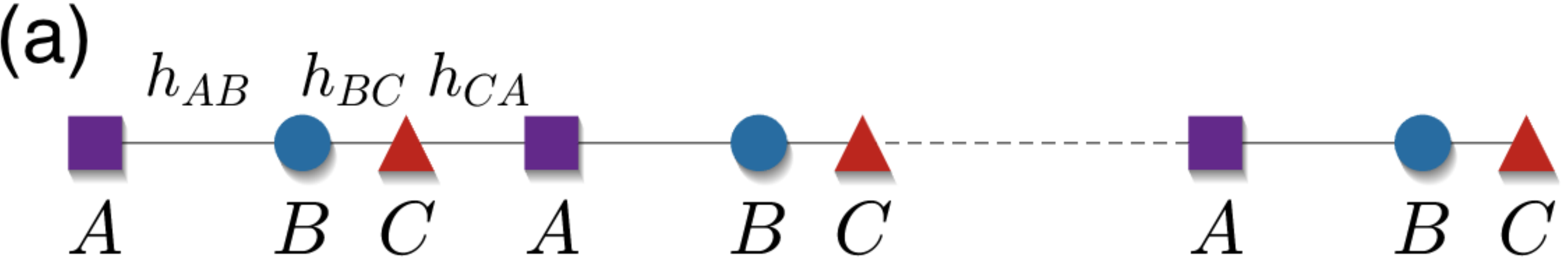}
\par\end{flushleft}
\includegraphics[width=1\textwidth]{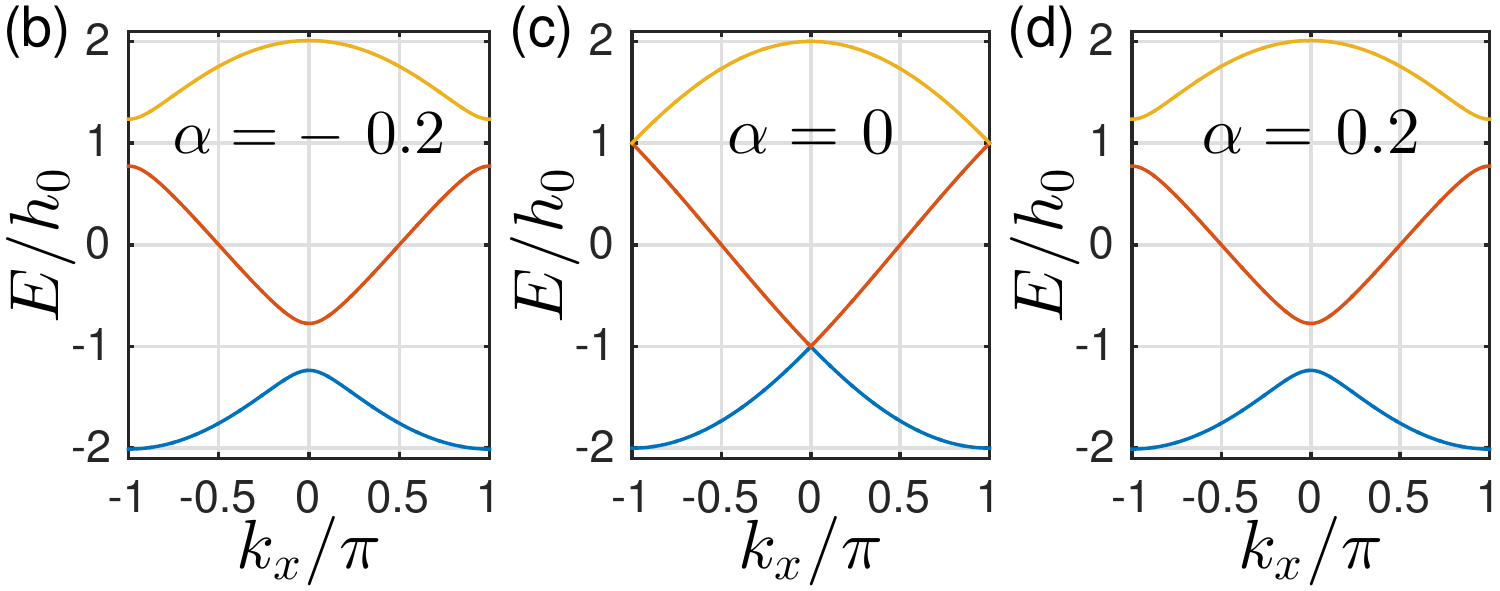}

\includegraphics[width=1\textwidth]{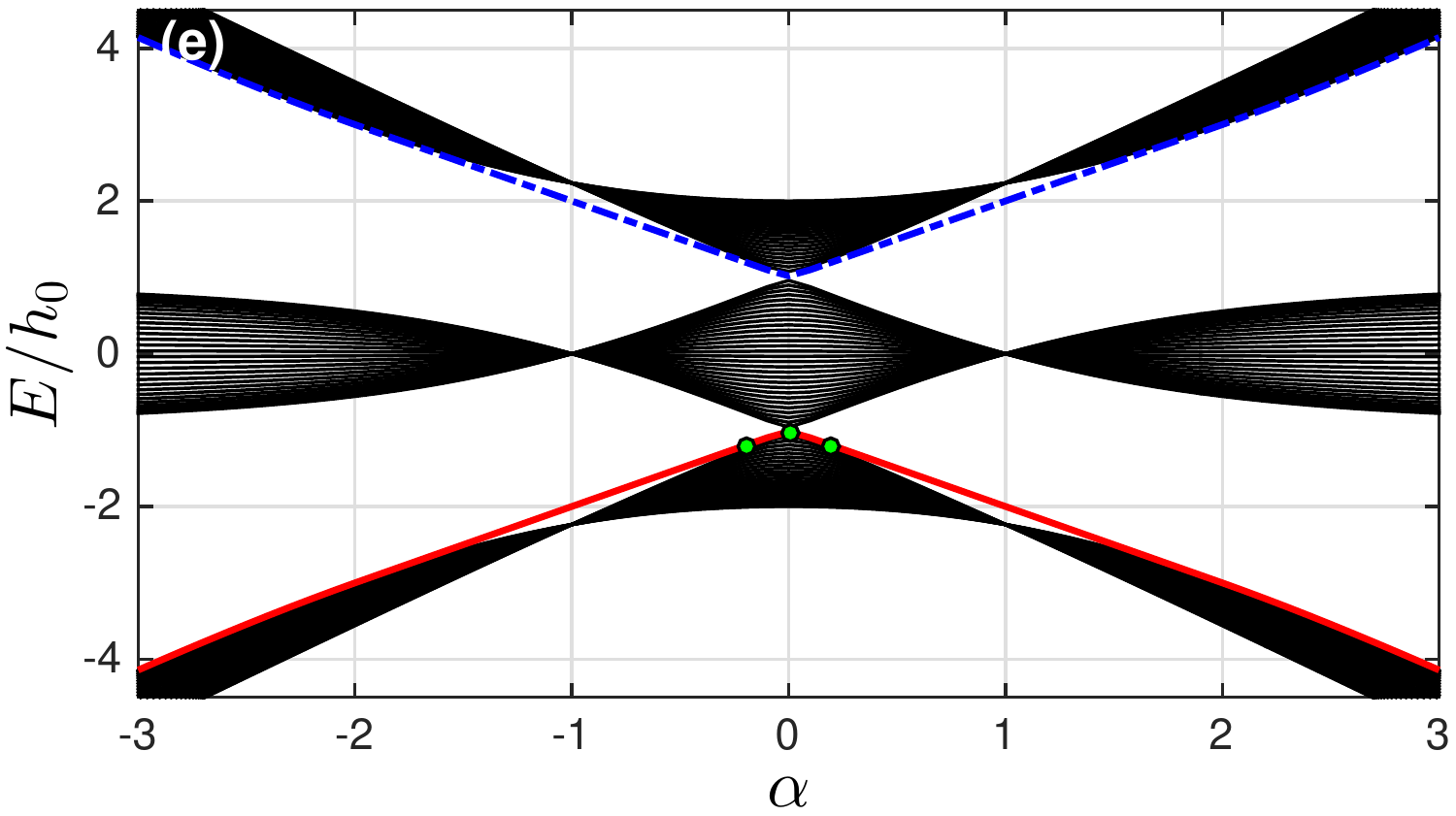}

\includegraphics[width=1\textwidth]{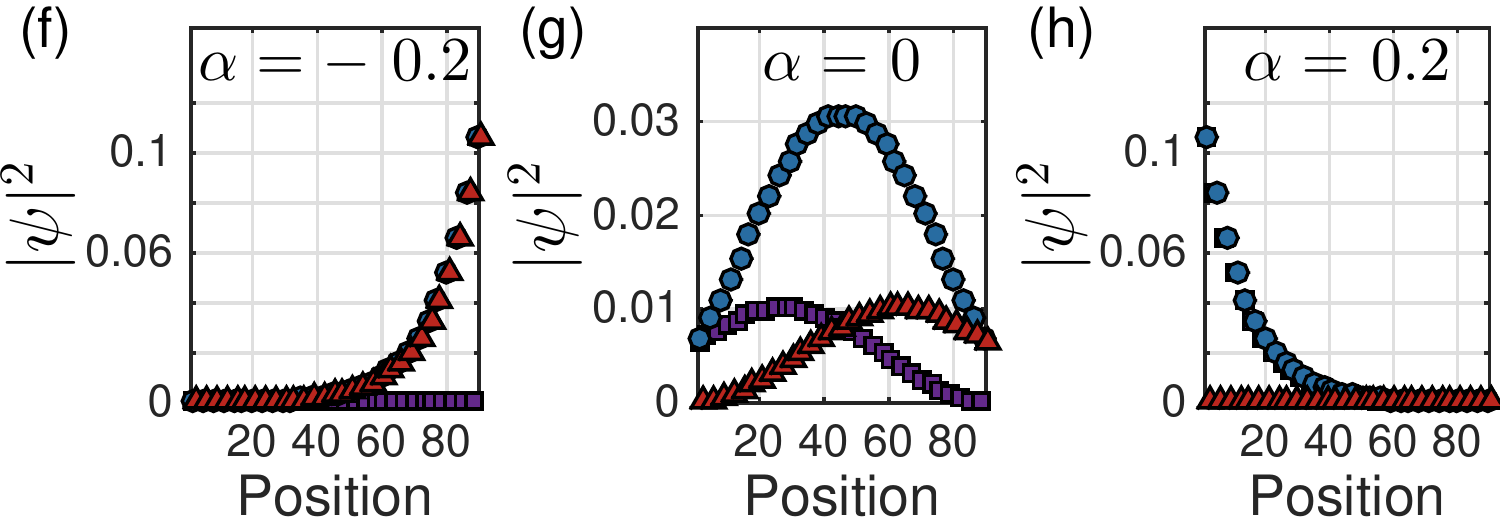}%
\end{minipage}\hfill{}%
\begin{minipage}[b]{0.42\textwidth}%
\includegraphics[width=1\textwidth]{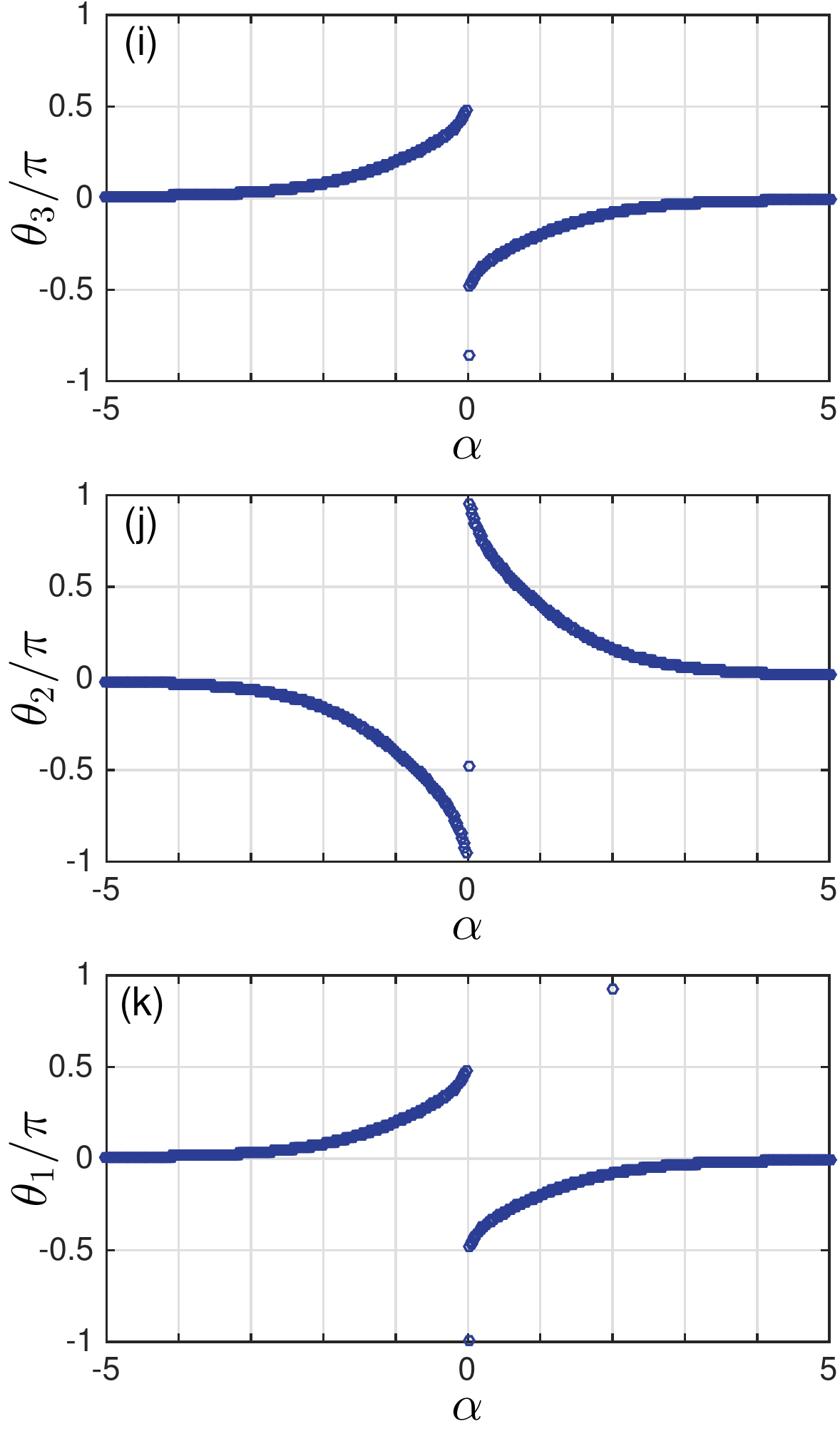}%
\end{minipage}%
\end{minipage}\caption{\label{fig1}(a), schematic picture of trimer lattices, each unit
cell contains three atoms. (b)-(d), spectrum of $\mathcal{H}\left(1,\alpha,k_{x}\right)$
(see Eq. \ref{eq:3}) as the function of $k_{x}/\pi$. (e), spectrum
of the corresponding finite trimer lattice $H\left(1,\alpha,W\right)$
as a function of $\alpha$. The width of sample is $W=30$ complete
unit cells, a.k.a. $90$ sites. Red solid line and blue dashed line
show the edge modes. The parts (f)-(h) gives the corresponding occupation
probabilities at the three green dots of the subplot \ref{fig1}(e):
purple square, distribution at sites $A$; blue dot, distribution
at sites $B$; red triangle, distribution at sites $C$. The parts
(i)-(k) gives the Berry phases of the three energy bands as the function
of $\alpha$. (i), the top band; (j), middle band; (k), the bottom
band.}
\end{figure*}

Thus to summarize the most important aspects of the 1D and the 2D
topology are the Berry Phase connection to the topological numbers
and the adiabatic pumping. In this article we present our theoretical
results on trimer 1D lattices, which can be used to demonstrate all
the new aspects of 1D phases. The photonic realization of the trimer
lattices is with in the current reach where waveguides can be written
on a chip by using femtosecond lasers \cite{Nicolo2012,Chaboyer2015,Robert2015}.
Further the bending of waveguides can be used to bring additional
dimensionality to the system and thus the new aspects of the 2D phases
can be studied. We discuss the new phases that can arise due to the
existence of a symmetry different from crystal symmetry, we call it
unit-cell symmetry (UCS). Our key findings are\textemdash{} 1. the
existence of piecewise continuous Berry phases which define two new
1D phases with the phase transition occurring at the discontinuity;
2. existence of edge modes localized at the opposite edges for the
two different phases and the tomography of such modes; 3. The 2D realization
using 1D lattice of trimer leads to phases characterized by very specific
2D Berry phases of half period, these characteristic Berry phases
change smoothly within a phase while change discontinuously at the
transition point; 4. The existence of adiabatic pumping for each phase;
5. Existence of gain assisted enhanced pumping; 6. Non-reciprocity
of the pumping process making the system a good optical diode. The
origin of non-reciprocity in our linear device is traced to certain
symmetry properties. This is distinct from recent apparatus based
on nonlinear optical methods \cite{XL2014,zongfu2009,chunhuadong2014,JunHwan2015,Fan2012,Ganainy2013}.
The addition of gain and loss is especially important for utilizing
edge modes for adiabatic pumping and the nonreciprocal behavior of
the system.

\section{New phases of 1D system: Piecewise Berry phase}

Our investigations are based on a 1D trimer lattice where each unit
cell consists of three sites with a specific form of symmetry to be
referred to as unit-cell symmetry (UCS in short). The fundamental
eigen equations for a trimer lattice are given by
\begin{align}
\varepsilon\psi_{n,A} & =\varepsilon_{0,A}\psi_{n,A}+h_{AB}\psi_{n,B}+h_{CA}\psi_{n-1,C}\nonumber \\
\varepsilon\psi_{n,B} & =h_{AB}\psi_{n,A}+\varepsilon_{0,B}\psi_{n,B}+h_{BC}\psi_{n,C}\label{eq:1}\\
\varepsilon\psi_{n,C} & =h_{CA}\psi_{n+1,A}+h_{BC}\psi_{n,B}+\varepsilon_{0,C}\psi_{n,C},\nonumber
\end{align}
here $\varepsilon$ is the eigen energy, $n$ is the index of unit
cell, $A,B,C$ label the three different sites in each unit cell (Fig.
\ref{fig1}. very top part (a)), $h_{AB}$, $h_{BC}$, $h_{CA}$ are
the coupling between the two sites, they are real; $\varepsilon_{0,A}$,
$\varepsilon_{0,B}$, $\varepsilon_{0,C}$ are the on-site energies
which we assume to be real and equal. As non-zero values of the on
site energies give over-all energy shift, we can set these as zero.
However positive or negative imaginary parts of on-site energies are
used in the latter discussion. This inclusion produces important new
results specifically in the context of pumping. With the same on-site
energies, UCS is given by the constraint on the coupling $h_{CA}=\left(h_{AB}+h_{BC}\right)/2$.
We will discuss more about this symmetry later. Now we want to show
that UCS makes the Berry phase piecewise continuous and can bring
two new phases of matter.

\begin{figure*}[t]
\begin{minipage}[t]{0.95\textwidth}%
\begin{minipage}[b]{0.49\textwidth}%
\includegraphics[width=1\textwidth]{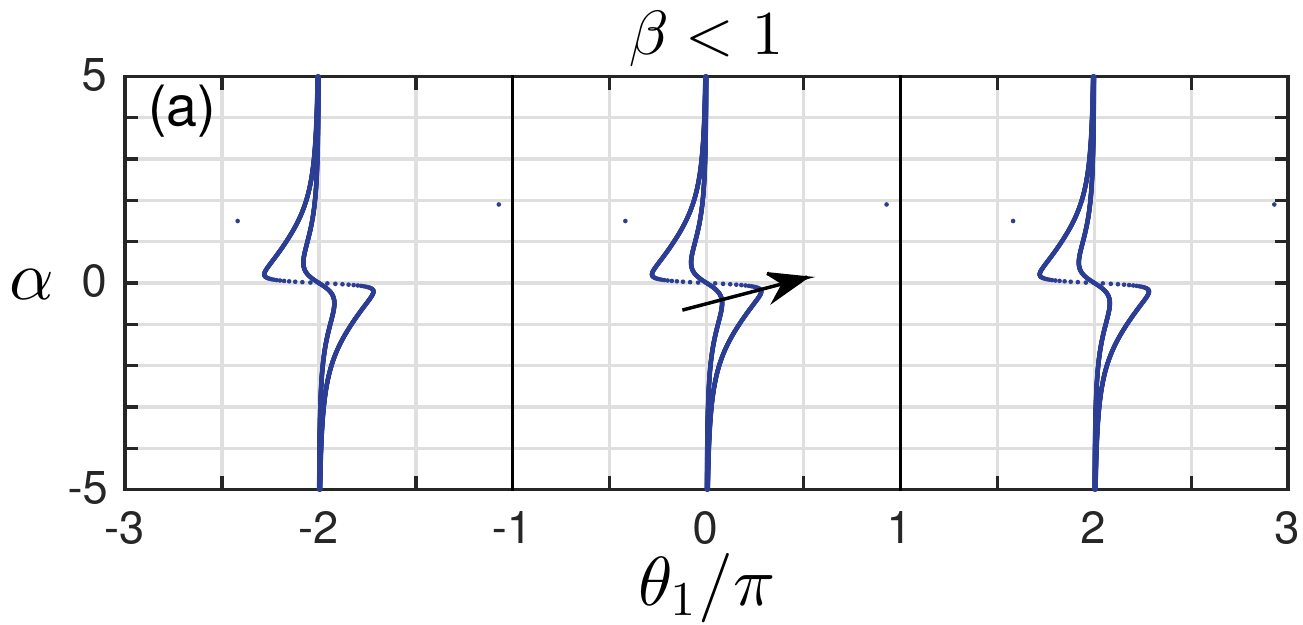}

\includegraphics[width=1\textwidth]{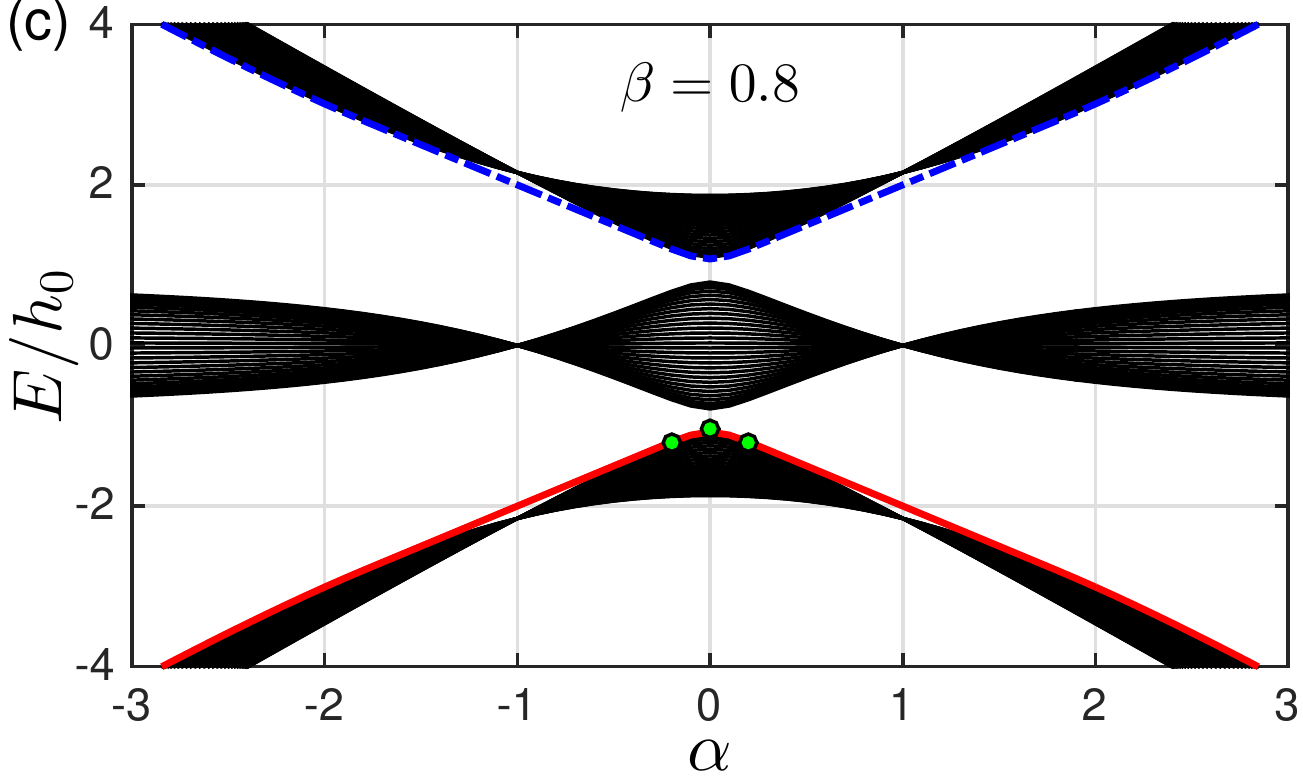}

\includegraphics[width=1\textwidth]{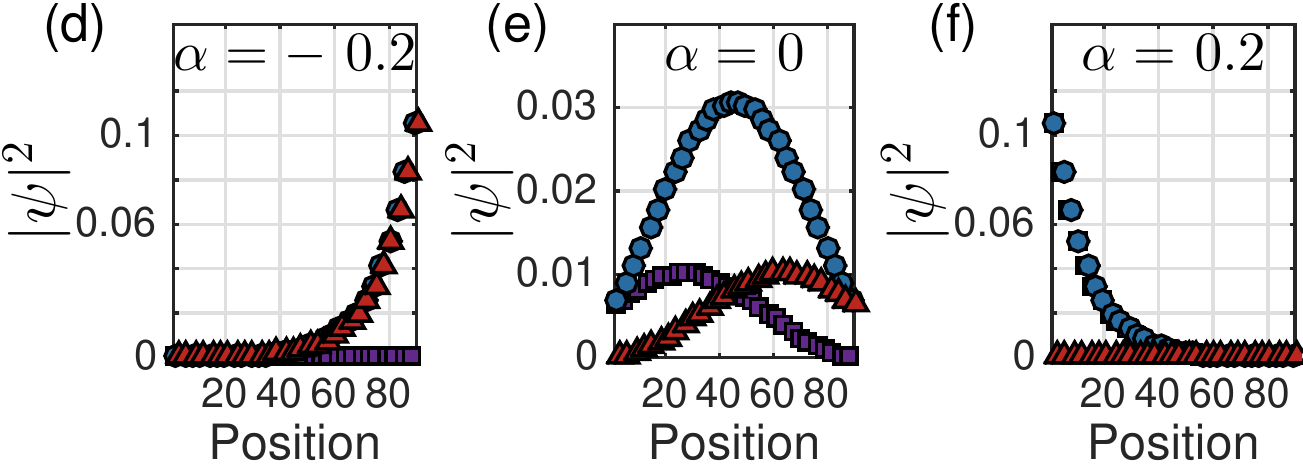}%
\end{minipage}\hfill{}%
\begin{minipage}[b]{0.49\textwidth}%
\includegraphics[width=1\textwidth]{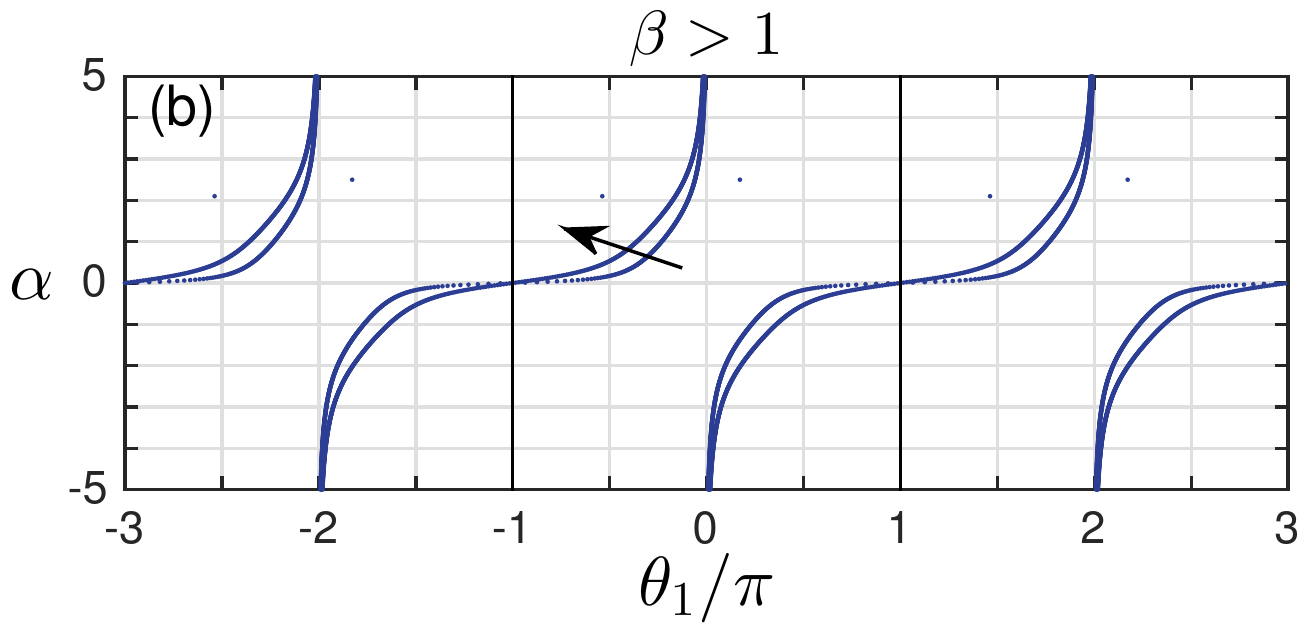}

\includegraphics[width=1\textwidth]{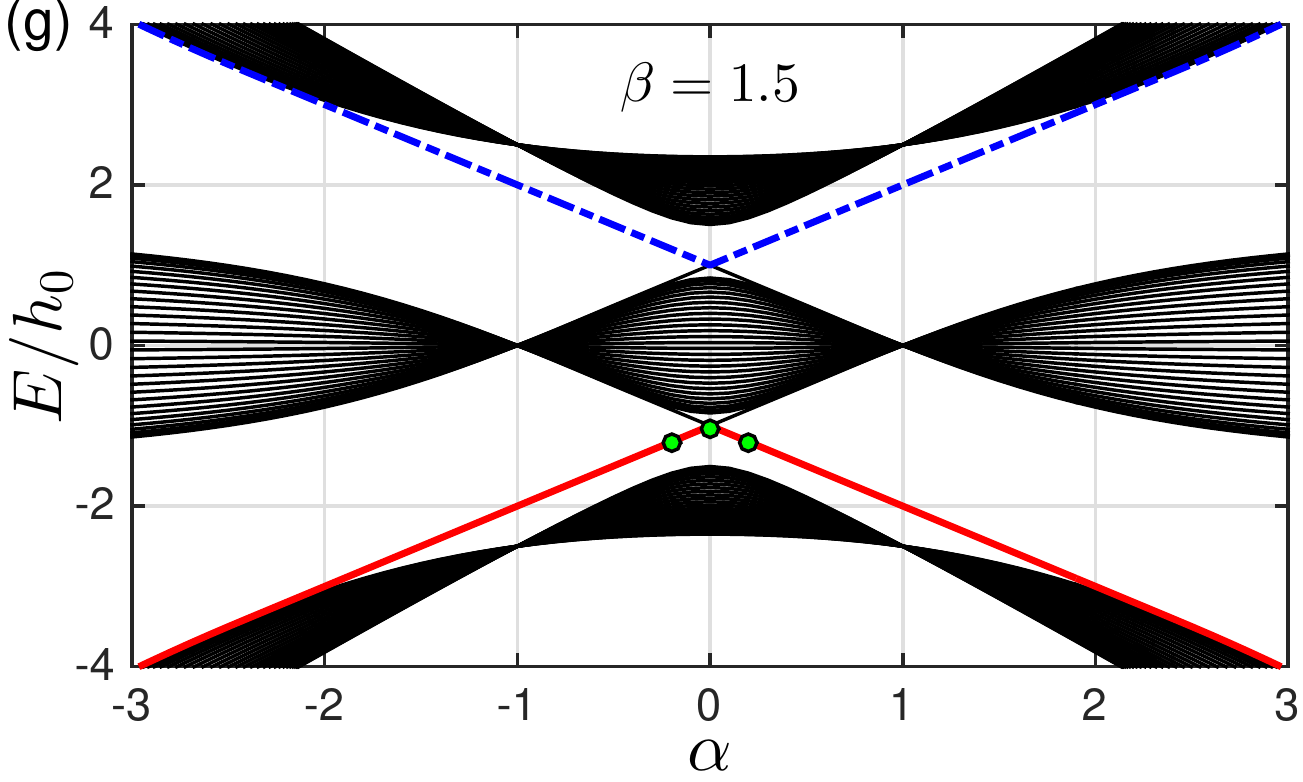}

\includegraphics[width=1\textwidth]{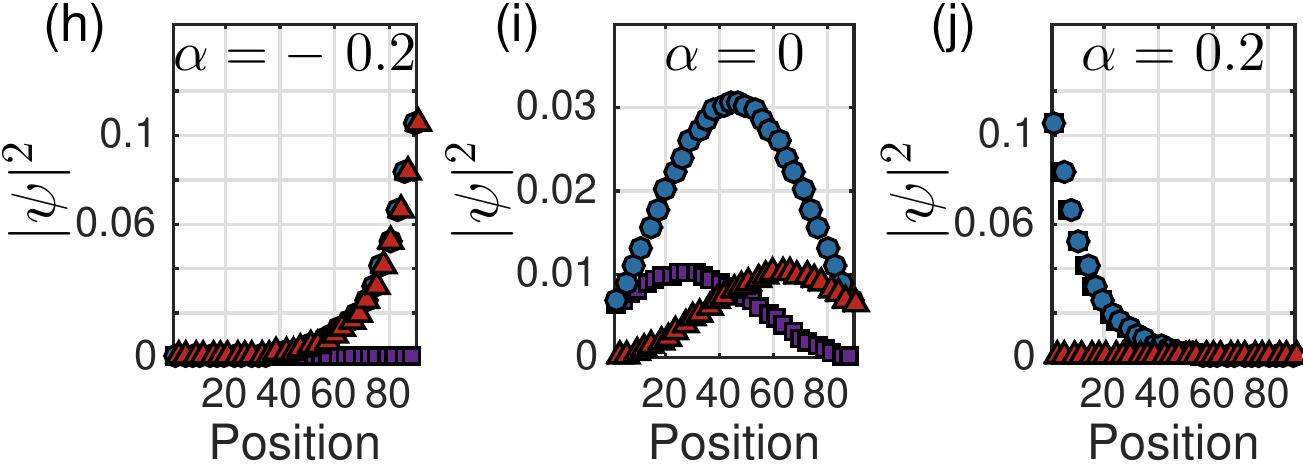}%
\end{minipage}%
\end{minipage}

\caption{\label{fig2}Berry phases, spectrum and eigen states of $\mathcal{H}\left(\beta,\alpha,k_{x}\right)$
(Eq. (\ref{eq:3})) for $\beta\protect\neq1$: left plots: $\beta<1$,
right plots: $\beta>1$. Plots (a)-(b) give evolution of Wannier center
(Berry phase) $\theta_{1}/\pi$ of the lowest band while $\alpha$
changes, three unit cells are shown (Compare to Fig.\ref{fig1} (i)-(k),
the x,y-coordinates are exchanged). Each plot contains two different
$\beta$, along the direction of arrow: (a). $\beta=0.5,0.9$; (b)
$\beta=1.1,1.5$. The plots (c),(g) show spectrum of finite trimer
lattice $H\left(\beta,\alpha,W\right)$ as the function of $\alpha$
for different $\beta$. The width of sample is $W=30$ unit cells.
The parts (d)-(f) and (h)-(j) give the corresponding occupation probabilities
at the position of three green dots of the two subplots (c),(g) separately.}
\end{figure*}

For the trimer lattice of infinite length, the system is translationally
invariant and can be described by the Bloch Hamiltonian
\begin{equation}
\mathcal{H}\left(k_{x}\right)=\left[\begin{array}{ccc}
\varepsilon_{0,A} & h_{AB} & h_{CA}e^{-ik_{x}}\\
h_{AB} & \varepsilon_{0,B} & h_{BC}\\
h_{CA}e^{ik_{x}} & h_{BC} & \varepsilon_{0,C}
\end{array}\right].\label{eq:2}
\end{equation}
The Berry phase of the 1D trimer lattice can be calculated by this
translationally invariant Hamiltonian. With $\varepsilon_{0,A}=\varepsilon_{0,B}=\varepsilon_{0,C}=0$,
by writing $h_{AB}=\bar{h}+h_{d}$ and $h_{BC}=\bar{h}-h_{d}$, the
Hamiltonian can be simplified as
\begin{equation}
\mathcal{H}\left(\beta,\alpha,k_{x}\right)=\left[\begin{array}{ccc}
0 & 1+\alpha & \beta e^{-ik_{x}}\\
1+\alpha & 0 & 1-\alpha\\
\beta e^{ik_{x}} & 1-\alpha & 0
\end{array}\right]\bar{h},\label{eq:3}
\end{equation}
where $\alpha=h_{d}/\bar{h}$, and the constraint $h_{CA}=\left(h_{AB}+h_{BC}\right)/2$
is given by $\beta=h_{CA}/\bar{h}=1$. It is convenient to choose
the overall factor $\bar{h}=1$. For the system has UCS, the eigen
problem of $\mathcal{H}\left(\beta=1,\alpha,k_{x}\right)$ can be
solved numerically. For each eigenstate $u_{n,k_{x}}$ $(n=1,2,3)$,
we can obtain the Berry-Wilczek-Zee connection $A_{n}\left(k_{x}\right)=-\textrm{i}\left\langle u_{n,k_{x}}\right|\frac{\partial}{\partial k}\left|u_{n,k_{x}}\right\rangle $
\cite{Berry1984,WilczekZee1984}, and calculate the corresponding
Berry phase $\theta_{n}=\int_{0}^{2\pi}A_{n}\left(k_{x}\right)dk_{x}$.
By setting $\theta_{i}\in\left[-\pi,\pi\right]$, the freedom of the
Berry phase is fixed, and has the physical meaning of center of mass
of electrons within one cell.

All the numerical results are shown in Figure \ref{fig1}. The parts
(b-d) give the energy bands of $\mathcal{H}\left(\beta=1,\alpha,k_{x}\right)$.
The band gaps are closed at $\alpha=0$ and opened when $\alpha>0$
or $\alpha<0$. The Fig. \ref{fig1}(e) gives the spectrum of trimer
lattice of finite length $W$ (see Eq. (1) for the Hamiltonian $H\left(\beta=1,\alpha,W\right)$),
i.e. the translational invariance is broken and $k_{x}$ is no more
a good quantum number. The edge modes occur for the finite trimer
lattice, they are localized at the left edge for $\alpha>0$ (Fig.
\ref{fig1}.(f)) and at the right edge for $\alpha<0$ (Fig.\ref{fig1}.(h)).

From the distinct behavior of eigenvalues and eigenstates, it is clear
that $\alpha>0$ and $\alpha<0$ are two different phases separated
by $\alpha=0$. We will now characterize these phases via the Berry
phases of the eigenfunctions of $\mathcal{H}\left(\beta=1,\alpha,k_{x}\right)$.
We can see that Berry phases of all the three bands are discontinuous
at $\alpha=0$ (Fig. \ref{fig1}. (i)-(k)). For the lowest band and
the top band, we have $\theta=\pi/2$ at $\alpha\rightarrow0^{+}$
and $\theta=-\pi/2$ at $\alpha\rightarrow0^{-}$. The Berry phase
$\theta$ decreases when $\left|\alpha\right|$ increases. For the
multi-electron ground state that only the lowest band is filled, this
picture means when the system is at the ground state, it is positively
polarized for $\alpha>0$ and negatively polarized for $\alpha<0$
, and the polarization reaches maximum at $\left|\alpha\right|\rightarrow0$.
Thus these two are physically distinct phases and disconnect with
each other unless the band gap is closing.

Note that $\alpha=0$, $\beta=1$ means that the hopping terms between
any two sites are same. The disconnection of Berry phase means $\alpha=0$
is not a stable. Any small disorder may make the lattice positively
polarized or negatively polarized. The instability at $\alpha=0$
is in fact the instability of the 1D lattice that is composed of the
same sites. The instability is the reason that 1D lattice can be dimerized
and described by the SSH model if the density of free electrons is
$1/2$ (i.e. on average every two sites contain one electron) \cite{Heeger1988}.
Our calculation shows that the 1D lattice is also not stable if the
electron density is $1/3$. When the system is $1/3$ filled, a small
gap is easily opened at the Fermi surface due to the instability,
which makes the ground state one stable trimerized phase. The two
phases of trimer lattice are the correspondence of the two phases
of dimerized lattice, we may call them $\alpha^{+}$ phase (for $\alpha>0$)
and $\alpha^{-}$ phase (for $\alpha<0$).

Now we may discuss the symmetry of the system. It is easy to check
that the Hamiltonian (\ref{eq:3}) has the inversion symmetry $U\mathcal{H}\left(\beta,\alpha,k_{x}\right)U^{-1}=\mathcal{H}\left(\beta,-\alpha,-k_{x}\right)$,
with the inversion matrix $U=\left[\begin{array}{ccc}
&  & 1\\
& 1\\
1
\end{array}\right]$. As Berry phase is the center of mass of electrons, the inversion
symmetry means center of mass of electrons is also inverted, i.e.
$\theta\left(\alpha\right)=-\theta\left(-\alpha\right)$. However,
this symmetry can not guarantee the discontinuity at $\alpha=0$.
As mentioned, the discontinuity is due to the UCS $h_{CA}=\left(h_{AB}+h_{BC}\right)/2$
or $\beta=1$. This can be seen from Fig. \ref{fig2}. The parts (a-b)
clearly show that for both $\beta>1$ and $\beta<1$, the Berry phases
are continuous at $\alpha=0$. Compare them to Fig. \ref{fig1}.(i)-(k),
it is clear that UCS $\beta=1$ guarantees two distinct phases for
$\alpha>0$ and $\alpha<0$. When UCS $\beta=1$ is broken, the system
can be continuously tuned from the $\alpha^{+}$ phase to the $\alpha^{-}$
phase without closing the band gap (see Figure \ref{fig2}. (c),(g)).
Thus the constraint $\beta=1$ has the same role as the symmetry on
the non-trivial topological system, though no global unitary symmetry
matrix $U\mathcal{H}\left(k_{x}\right)U^{-1}=\mathcal{H}\left(k_{x}\right)$
can be found and there is no correspondence with the crystal symmetry.
Since it is the constraint on the unit-cell, we call this constraint
as unit-cell symmetry (UCS).

\begin{figure}
\begin{centering}
\includegraphics[width=1\columnwidth]{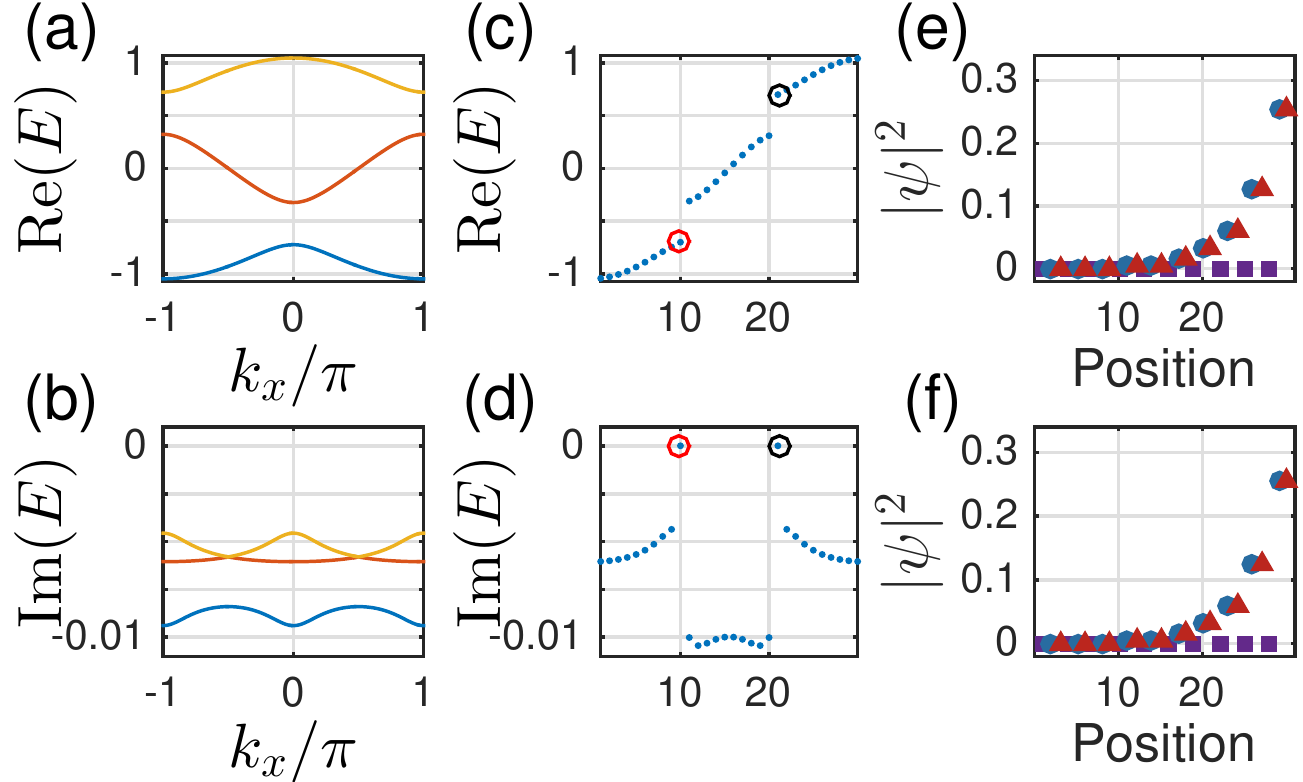}
\par\end{centering}
\caption{\label{fig3} (a), (b): for infinite complex trimer lattice, real
and imaginary spectrum as the function of dimentionless wavenumber
$k_{x}/\pi$; (c), (d): real and imaginary spectrum of a finite trimer
lattice which contains $W=10$ unit cells. The x-coordinates are the
index of the energies, which are ordered by the real parts. Two extra
edge modes are marked by red ($E_{1}$) and blue ($E_{2}$) circles;
(e), (f): distributions of $E_{1}$, $E_{2}$. }
\end{figure}

We next return to the question of edge and bulk modes for the model,
we will show the edge modes are robust and distinct from bulk modes
even when the system is open to the environment. For Eq. (\ref{eq:1}),
we choose the coupling between sites $h_{AB}=0.35h_{0}$, $h_{BC}=0.7h_{0}$
and $h_{CA}=0.5h_{0}$, i.e. $\alpha=-1/3$ and $\beta=1/1.05\simeq0.95$
in Eq. (\ref{eq:3}). Thus we look at the $\alpha^{-}$ phase. In
addition, we add a small imaginary part to the on-site energies, specifically
we choose $\varepsilon_{0,A}=-0.02h_{0}\textrm{i}$, $\varepsilon_{0,B}=0.02h_{0}\textrm{i}$
and $\varepsilon_{0,C}=-0.02h_{0}\textrm{i}$, i.e. sites $A$ and
$C$ have the same loss, but site $B$ has gain. We match the loss
and gain rates. Here the overall scaling factor $h_{0}$ only scales
the eigen spectrum and has no effect on the distributions. We choose
it as $1$ in further discussion. The plots in Fig. \ref{fig3} show
a very remarkable result: the edge modes do not decay while the bulk
states decay. The spectrum for infinite lattice Eq. (\ref{eq:2})
is given by Fig. \ref{fig3}.(a)-(b). Compared to infinite lattice,
the trimer lattice of finite length (\ref{eq:1}) contains two extra
modes. The real parts of eigen energies of these two edge modes are
in the real gap between bands (Fig. \ref{fig3}. (c)). The imaginary
parts of eigen energies of edge modes are exactly zero (Fig. \ref{fig3}.(d)).
In contrast, the imaginary part of the normal eigen modes is always
smaller than zero. This can be clearly seen from Fig. \ref{fig3}.
(d) or by comparing with Fig. \ref{fig3}. (b). The distribution of
the two edge modes is almost same and is localized at the right edge
of the lattice (Fig. \ref{fig3}. (e), (f)). We may also find that
the distributions at sites $B$ and $C$ are same and there is almost
no distribution at site $A$. The distributions decay fast from the
edge, roughly at rate $\sim\left(\frac{h_{AB}}{h_{BC}}\right)^{W-n}c_{0}=\left(\frac{1}{2}\right)^{W-n}c_{0}$,
here $n$ is the index of the unit cell, $W$ is the total width (i.e.
total number of unit cells), $c_{0}$ is the probability at the right
most side. In this sense, the two modes are called edge modes, and
other modes are called bulk modes. For such a system the propagation
of light can be used to do tomography of the non-decaying edge modes
as shown latter in Fig. \ref{fig5}.(b). This is because the bulk
modes decay away. It should be mentioned, the zero decay of edge modes
is due to the way we choose the imaginary part of the on-site energies.
However, even if we choose the imaginary part in a different way so
that edge modes also decay, both the real and imaginary spectrum of
edge modes are still away from bulk modes and the distributions are
still localized, which make them physically distinct from bulk modes.

\section{New phases of the 2D system characterized by the piecewise half-period
2D Berry phase}

For the 1D system, Figure \ref{fig2} shows that the two phases $\alpha>0$
and $\alpha<0$ are no more distinct when $\beta\neq1$. It also shows
that, $\beta>1$ and $\beta<1$ are the two different ways to break
the UCS $\beta=1$. As the physical meaning of Berry phase $\theta_{1}/2\pi$
is the center of mass of electron within the unit cell, Figure \ref{fig2}.(a)-(b)
shows the average motion of electron while $\alpha$ changes, three
connected cells are shown \cite{Alexandra2014}. For $\beta\neq1$,
the two disconnected piecewise Berry phases (Fig. \ref{fig1}.(h)-(j))
are now smoothly connected at $\alpha=0$, which means the electrons
can smoothly move from positive position to the negative position.
However, the motions for $\beta>1$ and $\beta<1$ are totally different.
For $\beta<1$, two pieces of Berry phases are connected at $\theta_{1}=0$,
the electron can only moves within one cell (Fig. \ref{fig2}(a))
. For $\beta>1$, two pieces of Berry phases are connected at the
cell boundary $\theta_{1}=\pm\pi$, the electron can moves from one
cell to another (Fig. \ref{fig2}(b)).

We may effectively build 2D material by smoothly connecting the two
1D phases $\alpha>0$ and $\alpha<0$ with the Hamiltonian,
\begin{equation}
\mathcal{H}\left(\beta,k_{y},k_{x}\right)=\left[\begin{array}{ccc}
0 & 1+\alpha\left(k_{y}\right) & \beta e^{-ik_{x}}\\
1+\alpha\left(k_{y}\right) & 0 & 1-\alpha\left(k_{y}\right)\\
\beta e^{ik_{x}} & 1-\alpha\left(k_{y}\right) & 0
\end{array}\right]\bar{h},\label{eq:4}
\end{equation}
with the periodic term $\alpha\left(k_{y}\right)=\alpha_{0}\cos k_{y}$.
From the inversion symmetry of the 1D system (\ref{eq:3}), it is
easy to check that the Hamiltonian (\ref{eq:4}) has the inversion
symmetry $U\mathcal{H}\left(\beta,k_{y},k_{x}\right)U^{-1}=\mathcal{H}\left(\beta,\pi-k_{y},-k_{x}\right)$
and correspondingly $\theta\left(k_{y}\right)=-\theta\left(\pi-k_{y}\right)$.
The inversion-symmetric topological insulators have been well discussed
for even-band systems \cite{Taylor2011,Alexandra2014}. Here for triple-band
system, we show that the cases $\beta>1$ and $\beta<1$ are two distinct
2D phases although they are not the topological system in the usual
sense. The discussion in the following supposes that only the lowest
band is filled, thus the edge modes between the top and middle bands
do not participate.

\begin{figure*}[t]
\begin{centering}
\includegraphics[width=0.95\textwidth]{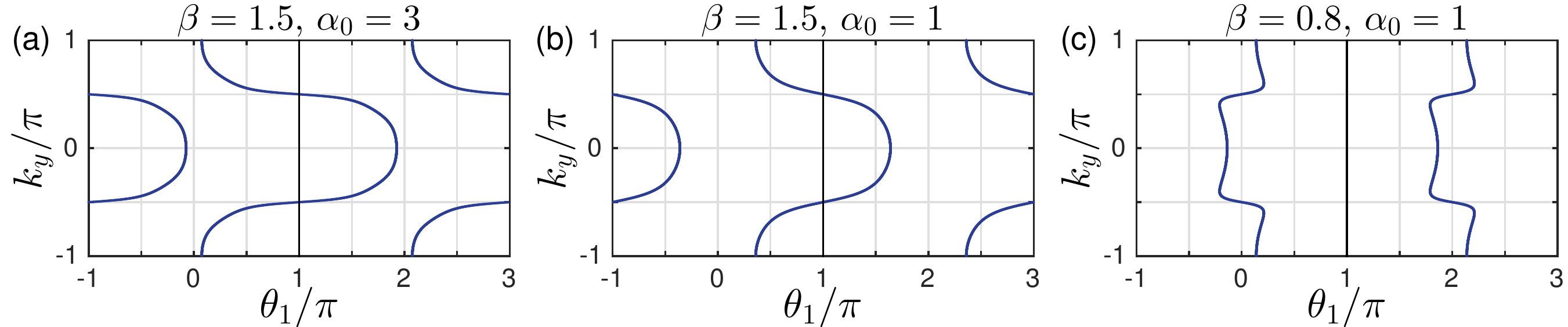}
\par\end{centering}
\begin{centering}
\includegraphics[width=0.95\textwidth]{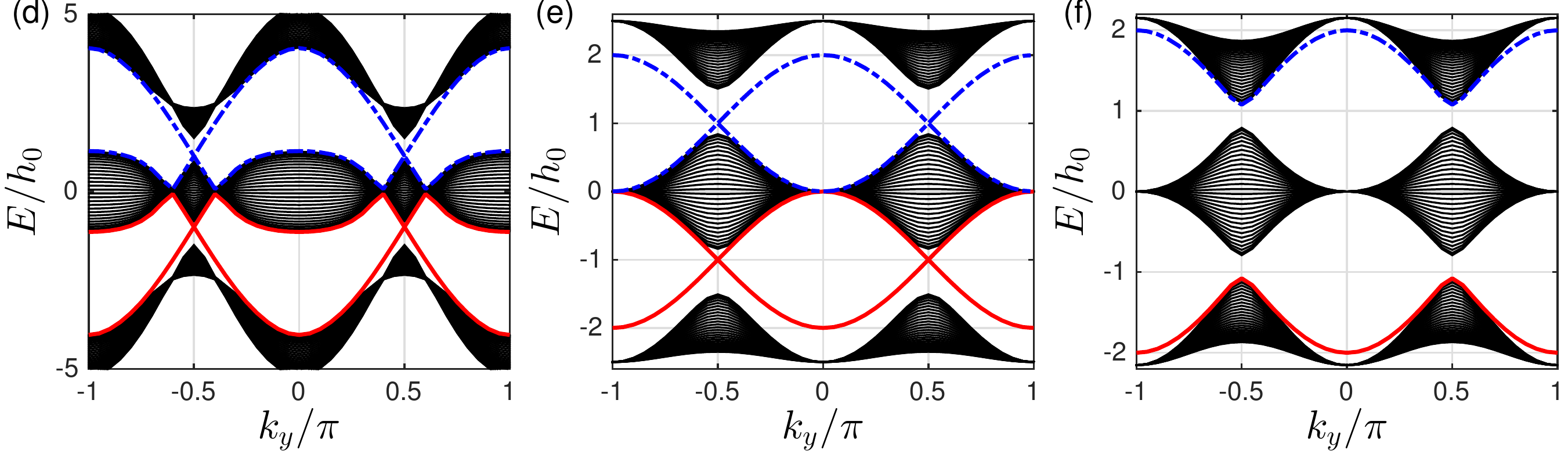}
\par\end{centering}
\caption{\label{fig4}(a)-(c): Wannier center flow of the lowest band of $\mathcal{H}\left(\beta,k_{y},k_{x}\right)$
(Eq.(\ref{eq:4})) for different $\alpha_{0}$ and $\beta$; (d)-(f):
The spectrum of the corresponding quasi-1D $H\left(\beta,k_{y}\right)$
as a function of $k_{y}$. The width of sample in $x$ direction is
$W=30$ unit cells. The sample is infinite in $y$ direction so that
$k_{y}$ is a good quantum number.}
\end{figure*}

For a translationally invariant system, non-zero integer numbers of
2D Berry phase $B_{2}=\frac{1}{2\pi}\intop_{k_{y}=-\pi}^{k_{y}=\pi}d\theta\left(k_{y}\right)$
are used to characterize the non-trivial 2D $\mathbb{Z}$ topology,
with $\theta\left(k_{y}\right)$ the effective 1D Berry phase of the
filled band for fixed $k_{y}$. However, for our model, we need some
other characteristic parameter to characterize phases. This is because
$B_{2}$ of Eq. (\ref{eq:4}), with the choice $\alpha\left(k_{y}\right)=\alpha_{0}\cos k_{y}$,
is always zero as $\theta_{1}\left(k_{y}\right)=\theta_{1}\left(-k_{y}\right)$,
the two half periods cancel each other (Fig. \ref{fig4}. (a-c)).
It is useful to introduce the Berry phase of half period ($\alpha\left(k_{y}\right)$
changes from $-\alpha_{0}$ to $\alpha_{0}$) $\tilde{B}_{2}=\frac{1}{2\pi}\intop_{k_{y}=-\pi}^{k_{y}=0}d\theta_{1}\left(k_{y}\right)$
which is nonzero.

Let us first examine the details for the case $\beta>1$. It is clear
that when $\left|\alpha_{0}\right|\rightarrow\infty$, the behavior
of the half period Berry phase like a Chern insulator, as $\tilde{B}_{2}$
is an integer. In particular, with the integer number $\tilde{B}_{2}=1$
as $\theta_{1}$ smoothly goes from $0$ to $2\pi$ in the half period
$-\pi\leq k_{y}\leq0$ and $-\alpha_{0}\leq\alpha\leq\alpha_{0}$
with $\left|\alpha_{0}\right|\rightarrow\infty$. This behavior can
be seen from Fig.\ref{fig2}. (b) for $\alpha_{0}\sim5$. Because
$\theta_{1}\left(k_{y}\right)$ interpolates across the maximal possible
range $\left[-\pi,\pi\right]$ within one $k_{y}$ cycle $-\pi\leq k_{y}\leq\pi$
\cite{Alexandra2014}, non-trivial 2D $\mathbb{Z}$ topology can be
defined for $\left|\alpha_{0}\right|\rightarrow\infty$. When the
system is finite in $x$-direction, Fig.\ref{fig2}. (g) shows that
the edge mode bands in the bulk-band gap smoothly connect the two
bulk bands. This connection makes the system 'gapless', and these
edge modes are called gapless edge mode, which are believed to be
the characteristic behavior of the non-trivial topology.

However, if $\left|\alpha_{0}\right|$ is finite, $\alpha_{0}\cos k_{y}$
changes from $-\alpha_{0}$ to $\alpha_{0}$, the tails of the 1D
Berry phase $\theta_{1}$ in Fig.\ref{fig2}. (b) for $\left|\alpha\right|>\left|\alpha_{0}\right|$
are not included in the integral of $\tilde{B}_{2}$, we should have
$\tilde{B}_{2}<1$. For one specific $\beta$, the tail is determined
by 1D Berry phase $\delta_{\alpha_{0}}\left(\beta\right)=\theta_{1}\left(\beta,-\alpha_{0}\right)=\theta_{1}\left(\beta,k_{y}=-\pi\right)$,
it can be shown that $0\leq\delta_{\alpha_{0}}\left(\beta\right)\leq\frac{\pi}{2}$
(see Fig. \ref{fig1}(j), Fig. \ref{fig2}(a)-(b) or Fig. \ref{fig4}(a)-(c)).
Due to the inversion symmetry, we may find $\tilde{B}_{2}=1-\frac{\delta_{\alpha_{0}}\left(\beta\right)}{\pi}$.
Thus for finite $\left|\alpha_{0}\right|$, the system is no more
topologically non-trivial as $\tilde{B}_{2}$ is no more an integer.
However, it is still possible to get the gapless edge modes which
connect the two bands when $\left|\alpha_{0}\right|$ is large enough
(Fig. \ref{fig4}. (d) with $\left|\alpha_{0}\right|=3$), thus the
occurrence of the gapless edge mode is not the characteristic behavior
of topological states of matter, it can occur even when $\tilde{B}_{2}$
is non-integer. On the other hand, the gapless edge modes are not
necessary for $\beta>1$. If $\left|\alpha_{0}\right|$ is small,
for example $\left|\alpha_{0}\right|=1$ (Fig. \ref{fig4}. (e)),
the edge modes will not connect to the bottom band. We can smoothly
change $\left|\alpha_{0}\right|$ from infinite to any finite value
without closing the gap (Fig.\ref{fig2}. (g)). This means that all
of them should belong to the same phase of $\beta>1$. Discrete characteristic
number, gapless edge modes are no more the signatures of the phase.
Instead, the characteristic behavior of the phase is that the edge
modes must connect to the middle band (Fig. \ref{fig4}. (d-e)). A
consequence of it is a measurable quantized Hall conductance. We can
also conclude that, the phase for $\beta>1$ implies that the the
center of mass of electron oscillates at the boundary of two unit
cells.

In contrast, for $\beta<1$, the center of mass of electron oscillates
around the center of unit cell (Fig. \ref{fig4}. (c)). The half period
Berry phase is given by $\tilde{B}_{2}=-\frac{\delta_{\alpha_{0}}\left(\beta\right)}{\pi}$
with $\delta_{\alpha_{0}}\left(\beta\right)=\theta_{1}\left(\beta,-\alpha_{0}\right)$.
Now, the edge modes are within the bottom band, and directly connect
to it at $\alpha_{0}\cos k_{y}=0$. Thus no edge modes occur in the
gap (Fig. \ref{fig4}. (f)), the Hall conductance purely due to the
edge modes is hard to get.

Another difference between the phases for $\beta>1$ and $\beta<1$
is the asymptotic behavior. For $\left|\alpha_{0}\right|\rightarrow\infty$,
the oscillation of center of mass of electron is pronounced for $\beta>1$
while is negligible for $\beta<1$. In contrast, for $\left|\alpha_{0}\right|\rightarrow0$,
the oscillation of center of mass of electron is relatively small
for $\beta>1$ while is pronounced for $\beta<1$.

It is clear that $\beta>1$ and $\beta<1$ are two distinct 2D insulator
phases for a fixed value of $\alpha_{0}$. The 2D phase is characterized
by the half period Berry phase $\tilde{B}_{2}=\frac{1}{2\pi}\intop_{k_{y}=-\pi}^{k_{y}=0}d\theta_{1}\left(k_{y}\right)$.
For $\beta=1$ (Fig. \ref{fig1}), within the first Brillouin zone,
the gaps of the two energy bands of the 2D material $\mathcal{H}\left(\beta,k_{y},k_{x}\right)$
(Eq. \ref{eq:4}) are closed at $k_{x}=0,2\pi$, and $\alpha=\alpha_{0}\cos k_{y}=0$,
i.e. $k_{y}=\pi/2,3\pi/2$. The gap closing witnesses a phase transition.
We find that $\tilde{B}_{2}$ jumps from $-\frac{\delta_{\alpha_{0}}\left(1\right)}{\pi}$
(for $\beta<1$) to $\tilde{B}_{2}=1-\frac{\delta_{\alpha_{0}}\left(1\right)}{\pi}$
(for $\beta>1$). Here $\delta_{\alpha_{0}}\left(1\right)=\theta_{1}\left(\beta=1,-\alpha_{0}\right)$
at the gap closing can be directly obtained from Fig. \ref{fig1}(k).

The two 2D phases $\beta>1$ and $\beta<1$ reflect the boundary physics
along $x$ direction. This is because the difference between the two
phases are the oscillation positions of the center of mass of electrons,
which depends on the choice of the unit cell along $x$ direction
\cite{Alexandra2014}. If we choose a new cell by a shift so that
the the center and the boundary are exchanged, and if the trimer lattice
consists of such cells then the physics of two phase is interchanged.

\begin{figure*}
\begin{minipage}[b]{0.465\textwidth}%
\includegraphics[width=1\textwidth]{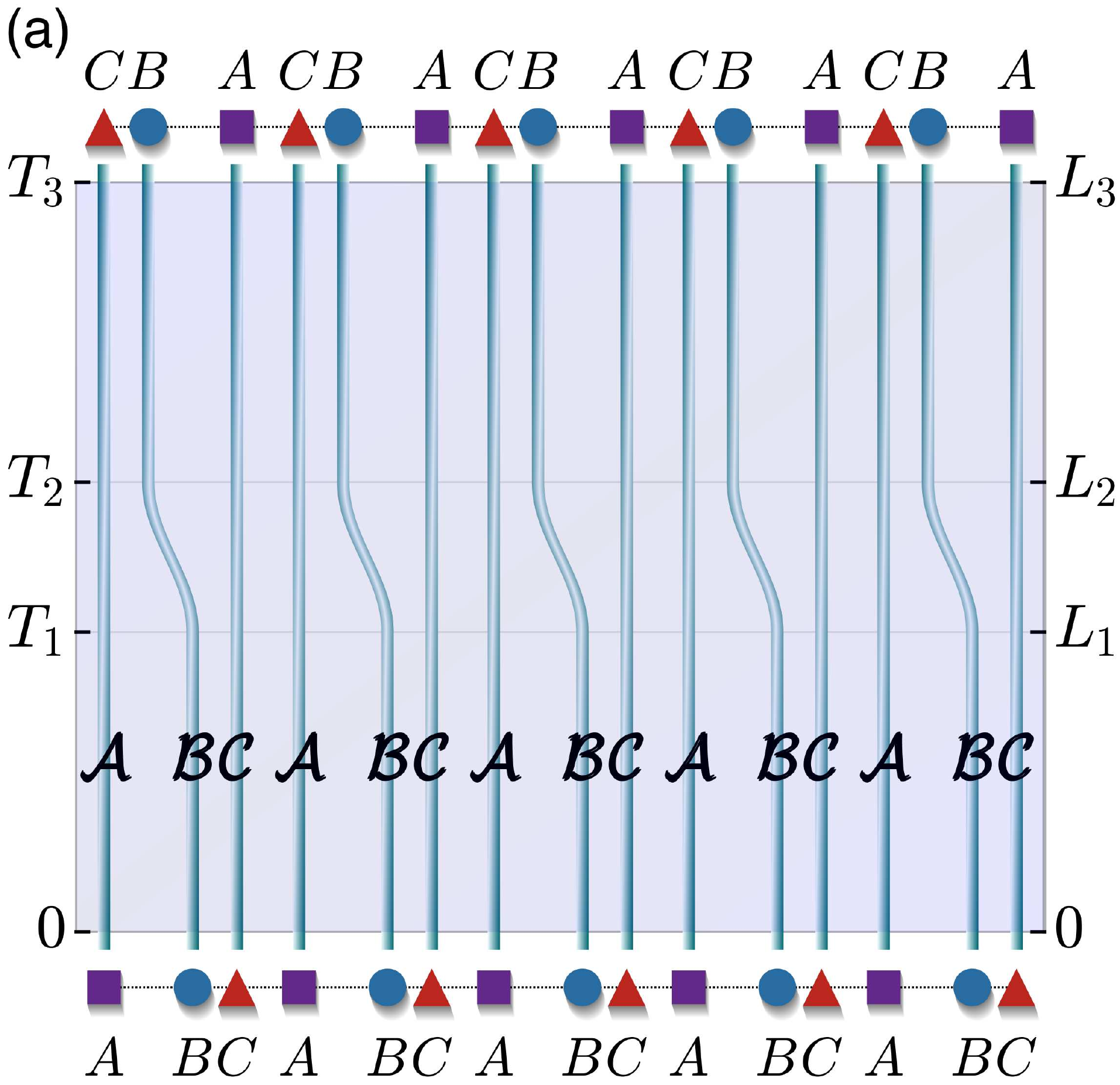}

\vspace{0.25cm}

\includegraphics[width=1\textwidth]{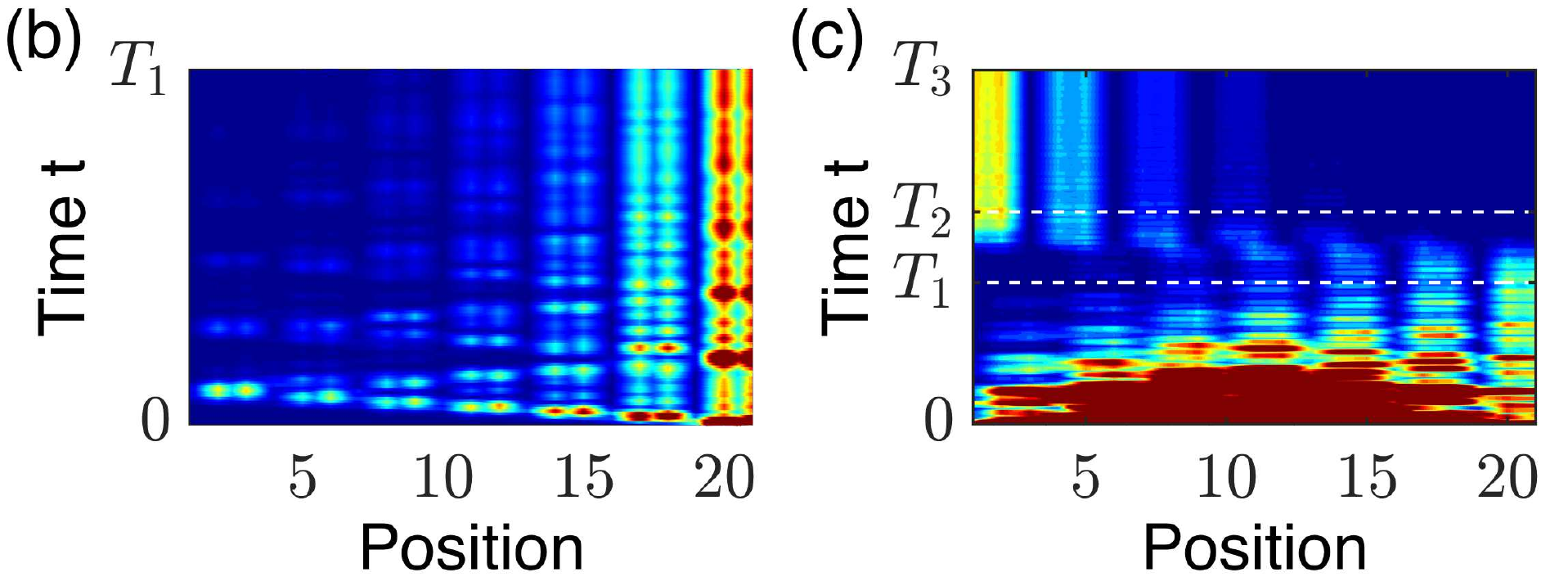}%
\end{minipage}\hfill{}%
\begin{minipage}[b]{0.515\textwidth}%
\includegraphics[width=1\textwidth]{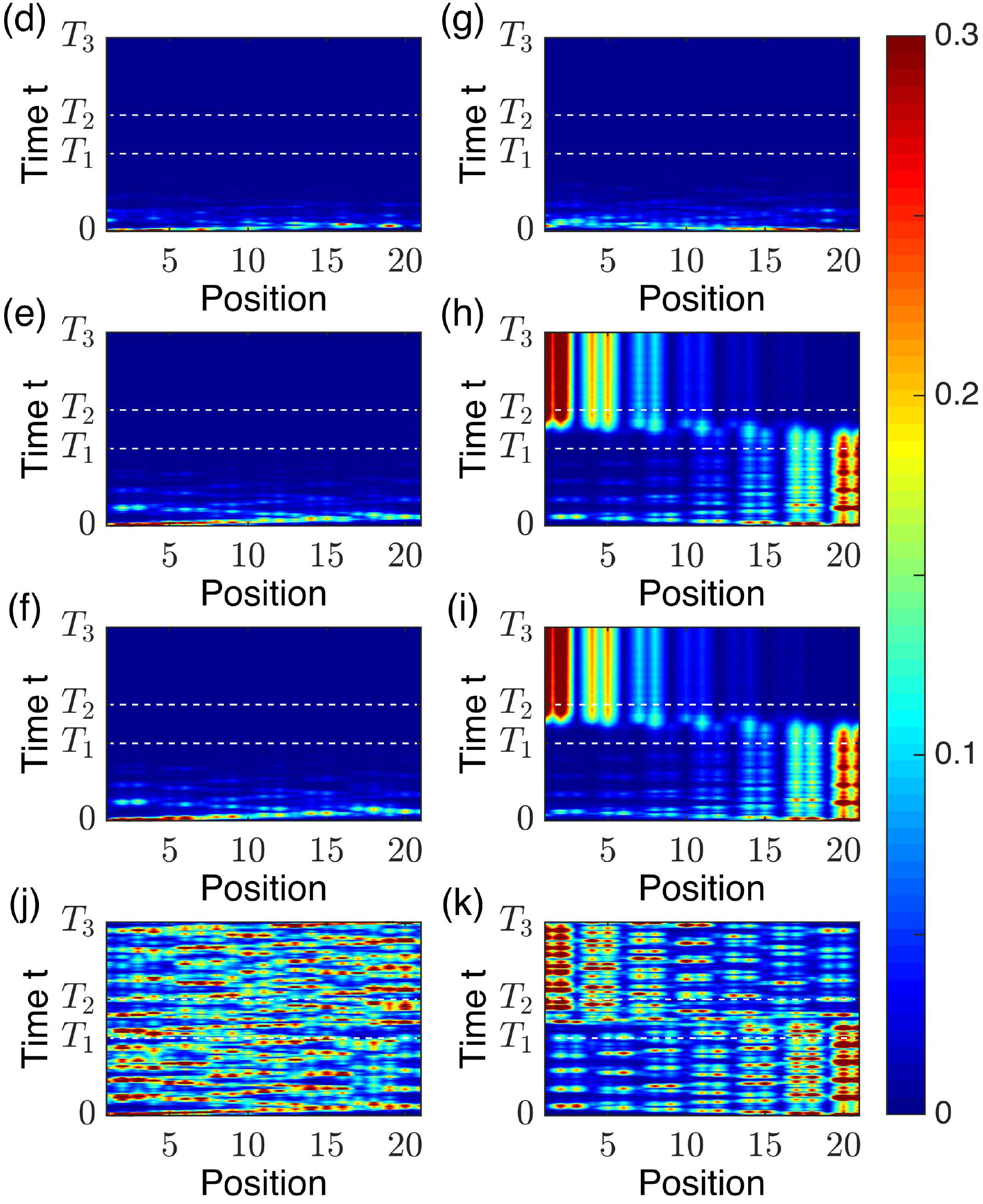}%
\end{minipage}

\caption{\label{fig5} (a). schematic picture of pumping by tuning the hopping
rates. The sites of the trimer lattice are marked by $A,B,C$, while
the sites of waveguides are marked by calligraphic symbol $\mathcal{A},\mathcal{B},\mathcal{C}$.
The bottom of waveguides is equivalent to the trimer lattice at $\alpha^{-}$
phase while the top is equivalent to the trimer lattice at $\alpha^{+}$
phase. The distance between waveguides is chosen as the inverse of
hopping, e.x., $l_{\mathcal{A}\mathcal{B}}=1/h_{\mathcal{A}\mathcal{B}}$.
(b-k), Evolution of lights along the tuned waveguides of Fig. \ref{fig5}.
(a) with different initial states. The width of the sample is $W=7$
unit cells ($21$ waveguides). Parameters of the time dependent Hamiltonian
$H\left(t\right)$ are given by Eq. (\ref{eq:6a},\ref{eq:6b}) and
the contexts. (b), Only the first time range $t\in\left[0,T_{1}\right]$
is considered, with $T_{1}=500/h_{0}$. We have the tomography of
edge modes, with the light fed in from the most right $\mathcal{B}$.
For the rest of subplots (c-k), the time ranges are chosen: $T_{1}=2T_{0}$,
$T_{2}=3T_{0}$ and $T_{3}=5T_{0}$ with $T_{0}=200/h_{0}$. (c),
Evenly feed lights from all the waveguides. (d-f), Feed in light from
the most left unit cell: (d). fed in from the most left site, i.e.
the $\mathcal{A}$ waveguide; (e)-(f), feed in from the most left
$\mathcal{B}$, $\mathcal{C}$ separately. (g-i), Fed in light from
the most right $\mathcal{A}$, $\mathcal{B}$, $\mathcal{C}$ waveguides
separately. (j-k), Without the imaginary part of on-site energies
$\varepsilon_{0,\mathcal{A}}=\varepsilon_{0,\mathcal{B}}=\varepsilon_{0,\mathcal{C}}=0$:
(j), from the most left wave guide $\mathcal{B}$; (k), from the most
right waveguide $\mathcal{B}$. }
\end{figure*}

\section{2D Lattice: the pumping process}

\begin{figure*}[t]
\begin{minipage}[b]{0.49\textwidth}%
\includegraphics[width=1\textwidth]{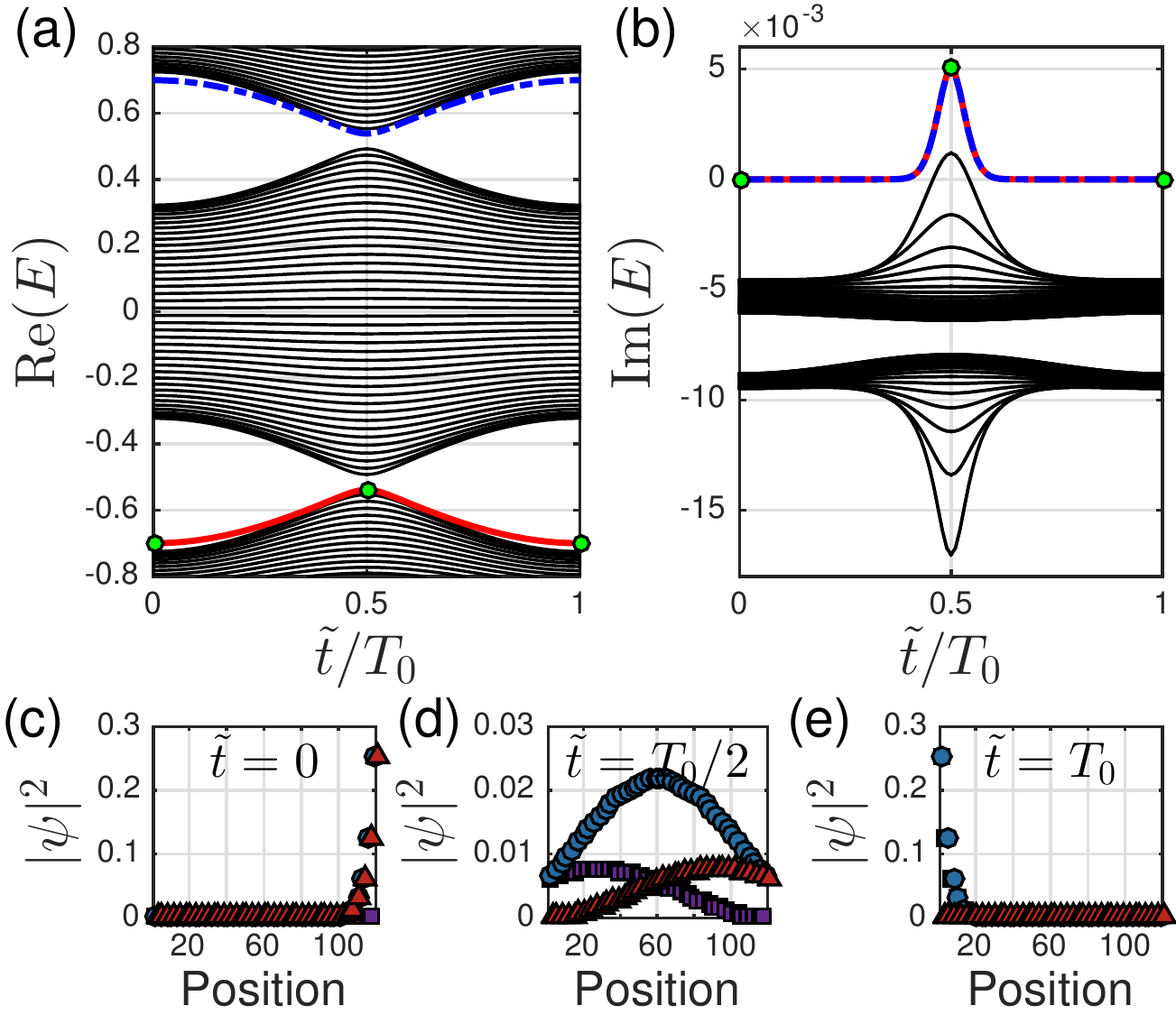}%
\end{minipage}\hfill{}%
\begin{minipage}[b]{0.49\textwidth}%
\includegraphics[width=1\textwidth]{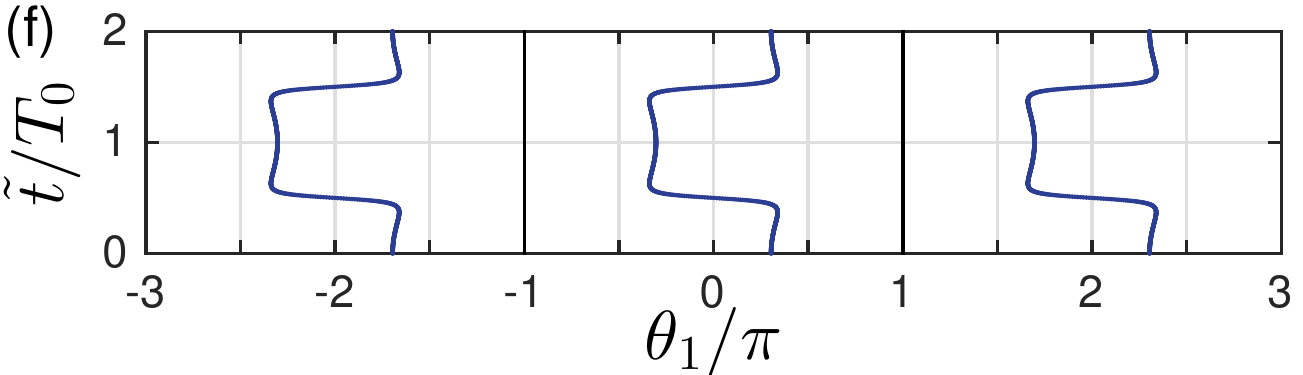}

\includegraphics[width=1\textwidth]{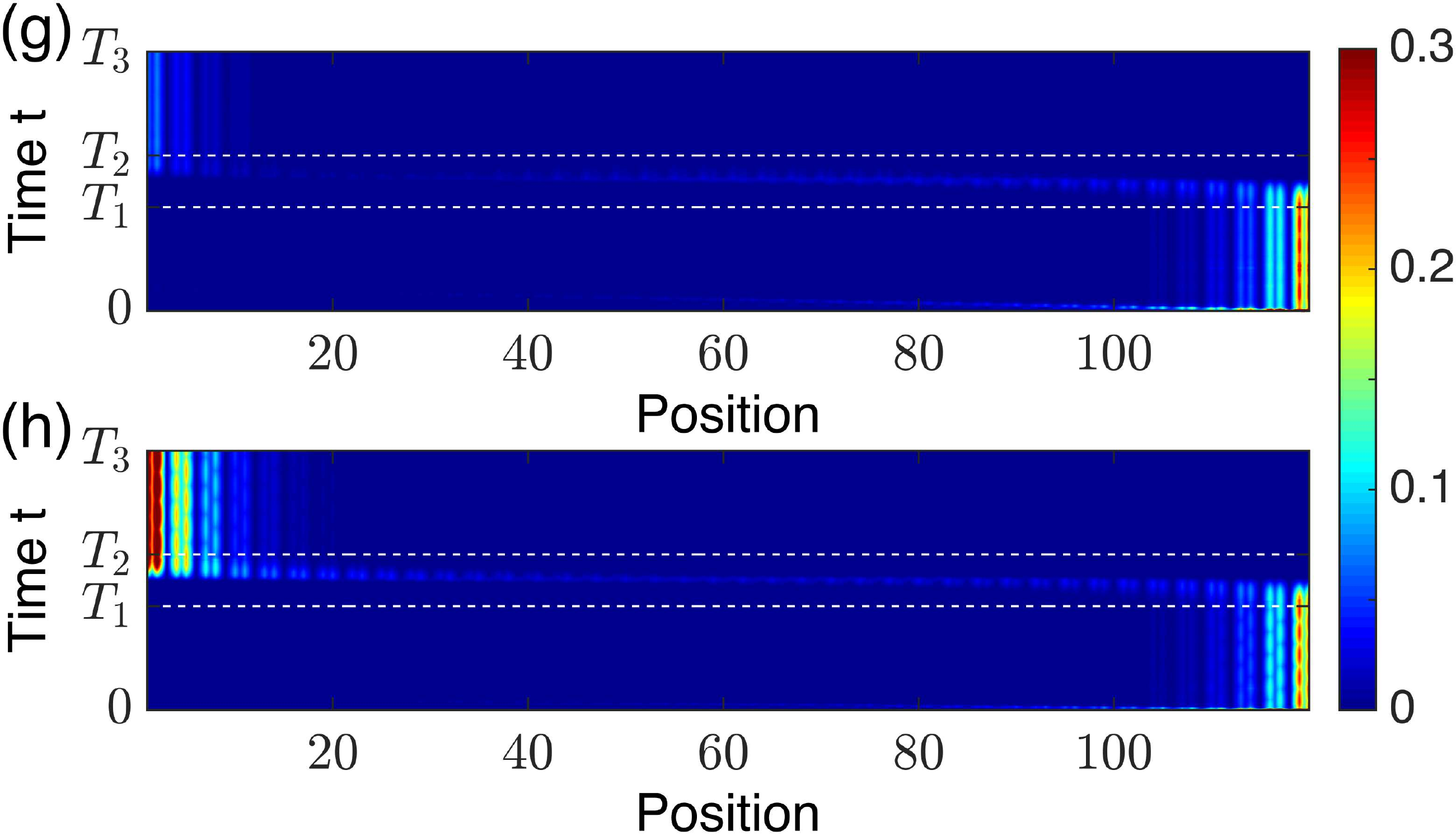}%
\end{minipage}

\caption{\label{fig6}(a)-(b), the real and imaginary parts of instantaneous
eigen energies $E\left(t\right)$ of the Hamiltonian $H\left(t\right)$
as a function of $\tilde{t}/T_{0}$ at the tuning range with $T_{0}=T_{2}-T_{1}$
and $\tilde{t}=t-T_{1}$. Only a half period $\frac{\pi\tilde{t}}{T_{0}}\in\left[0,\pi\right]$
is shown. Here we choose the width of the sample as $W=40$ unit cells.
Other parameters are given around Eq. (\ref{eq:6a}) and (\ref{eq:6b}).
The energies of the two 'edge' modes are marked by the blue dashed
line and the red solid line. (c-e) gives the distributions of the
red 'edge' modes at different time: (c), $\tilde{t}=0$ is corresponding
to $t=T_{1}$ of Fig. \ref{fig5} and (e), $\tilde{t}=T_{0}$ is corresponding
to $t=T_{2}$; (d), $\tilde{t}=T_{0}/2$ is at the middle of the tuning
range. (f). The corresponding change of Wannier center $\theta_{1}/\pi$
when $\tilde{t}/T_{0}$ changes. (g)-(h), enhanced pumping due to
the behavior around $T_{0}/2$. The lights are input from the most
right $\mathcal{C}$ waveguide. For $W=40$ unit cells, the tuning
time are (g). $T_{0}=600/h_{0}$; (h). $T_{0}=1200/h_{0}$.}
\end{figure*}

As mentioned in the beginning, non-trivial 2D topology is demonstrated
by adiabatic pumping. However, non-trivial topology is not the necessary
condition for pumping. For our system, both the 2D phases can give
the pumping process. This is especially realized in the context of
photonic lattices i.e. an array of waveguides as shown in Fig. \ref{fig5}.(a)\cite{Yaacov2012,H.Schomerus2013}
(See Appendix for more details). Light fed in from the bottom of waveguides
propagates in the three parts in the time range $t\in\left[0,T_{1}\right]$,
$\left[T_{1},T_{2}\right]$ and $\left[T_{2},T_{3}\right]$ separately.
The time-dependent Hamiltonian $H\left(t\right)$ (for finite number
of unit cell) or $\mathcal{H}\left(k_{x},t\right)$ (for infinite
number of unit cell) of light has the form (\ref{eq:1}) or (\ref{eq:2})
with the new parameters. For example, (\ref{eq:2}) is replaced by
\begin{equation}
\mathcal{H}\left(k_{x},t\right)=\left[\begin{array}{ccc}
\varepsilon_{0,\mathcal{A}} & h_{\mathcal{AB}}\left(t\right) & h_{\mathcal{CA}}e^{-ik_{x}}\\
h_{\mathcal{AB}}\left(t\right) & \varepsilon_{0,\mathcal{B}} & h_{\mathcal{B}\mathcal{C}}\left(t\right)\\
h_{\mathcal{CA}}e^{ik_{x}} & h_{\mathcal{B}\mathcal{C}}\left(t\right) & \varepsilon_{0,\mathcal{C}}
\end{array}\right].\label{eq:5}
\end{equation}
Now the site index $A,B,C$ is replaced by the waveguide index $\mathcal{A},\mathcal{B},\mathcal{C}$.
In the following, we check the pumping process for $\beta<1$ phase
for a half period. The parameters are based on the results of Fig.
\ref{fig3} for the $\alpha^{-}$ phase, which also gives decaying
bulk modes and satisfy the condition of easy tuning. For both $H\left(t\right)$
and $\mathcal{H}\left(k_{x},t\right)$, we set $\varepsilon_{0,\mathcal{A}}=-0.02h_{0}\textrm{i}$,
$\varepsilon_{0,\mathcal{B}}=0.02h_{0}\textrm{i}$, $\varepsilon_{0,\mathcal{C}}=-0.02h_{0}\textrm{i}$
so that the bulk states decay away in the first time range $t\in\left[0,T_{1}\right]$;
$h_{\mathcal{CA}}=0.5h_{0}$ is also fixed for the whole pumping process.
The time dependent parameters $h_{\mathcal{AB}}\left(t\right)$ and
$h_{\mathcal{BC}}\left(t\right)$ are piecewise function of $t$:
by setting $h_{AB}=0.35h_{0}$ and $h_{BC}=0.7h_{0}$, we have $h_{\mathcal{AB}}\left(t\right)\equiv h_{AB}$
and $h_{\mathcal{BC}}\left(t\right)\equiv h_{BC}$ in the range $t\in\left[0,T_{1}\right]$;
and $h_{\mathcal{AB}}\left(t\right)\equiv h_{BC}$ and $h_{\mathcal{BC}}\left(t\right)\equiv h_{AB}$
in the range $t\in\left[T_{2},T_{3}\right]$. The pumping process
in the range $t\in\left[T_{1},T_{2}\right]$ is modeling by a cosine
function, $h_{\mathcal{AB}}\left(t\right)$ and $h_{\mathcal{BC}}\left(t\right)$
are given by \begin{subequations}
\begin{alignat}{1}
h_{\mathcal{AB}}\left(t\right) & =\bar{h}+h_{d}\cos\frac{\pi\tilde{t}}{T_{0}},\label{eq:6a}\\
h_{\mathcal{BC}}\left(t\right) & =\bar{h}-h_{d}\cos\frac{\pi\tilde{t}}{T}.\label{eq:6b}
\end{alignat}
\end{subequations}Here $T_{0}=T_{2}-T_{1}$ and $\tilde{t}=t-T_{1}$.
The pumping range $t\in\left[T_{1},T_{2}\right]$ gives $\frac{\pi\tilde{t}}{T}\in\left[0,\pi\right]$,
i.e. only the half period is needed. We still set $\bar{h}=\frac{h_{AB}+h_{BC}}{2}$
and $h_{d}=\frac{h_{AB}-h_{BC}}{2}$. Obviously, $h_{\mathcal{AB}}\left(T_{1}\right)=h_{AB}$
and $h_{\mathcal{AB}}\left(T_{2}\right)=h_{BC}$, the Hamiltonian
of the three time range are connected. The parameters chosen above
also guarantee the band gaps of $\mathcal{H}\left(k_{x},t\right)$
or $H\left(t\right)$ are not closed for the whole pumping process
(Fig.\ref{fig6}.(a)-(b)). In this way we avoid the existence of the
exceptional points \cite{Demange2012} of the non-Hermitian hamiltonian,
and the system remains diagonalizable \cite{AABook2003}. However,
it can be check that pumping process may exist for the gap-closing
case.

The propagation of light in such a system can be studied by solving
the time dependent Schrödinger Equation $\textrm{i}\partial_{t}\Psi\left(t\right)=H\left(t\right)\Psi\left(t\right)$.
Numerically, we separate the pumping range $t\in\left[T_{1},T_{2}\right]$
in small time intervals, and suppose at each small interval $\left[t_{j},t_{j}+\triangle t\right]$,
the light propagates with the constant hamiltonian $H\left(t_{j}\right)$,
we then obtain\begin{subequations}
\begin{align}
\left|\Psi\left(t\right)\right\rangle  & \approx\prod_{j}^{N_{t}\leftarrow1}V\left(t_{j}\right)\textrm{e}^{-\textrm{i}D\left(t_{j}\right)\triangle t}V^{-1}\left(t_{j}\right)\left|\Psi\left(0\right)\right\rangle ,\label{eq:7a}\\
D\left(t\right) & =V^{-1}\left(t\right)H\left(t\right)V\left(t\right).\label{eq:7b}
\end{align}
\end{subequations}Thus $D\left(t\right)$ is the diagonalized form
of $H\left(t\right)$ for a given $t$, and the columns of $V\left(t\right)$
are the corresponding instantaneous eigenstates. Here $N_{t}$ is
number of time intervals, $\triangle t=T_{3}/N_{t}$ is the interval
length. $\prod_{j}^{N_{t}\leftarrow1}$ is the abbreviation of \textit{ordered}
matrix multiplication. As the matrices do not commute, the matrix
with smaller $j$ index should be at the right. As long as the time
interval $\triangle t$ is small enough, such a calculation is a good
approximation to the time-dependent Schrödinger Equation.

The evolution of lights for different initial states is shown in Fig.\ref{fig5}.
The initial states in the coordinates of waveguides can be easily
given. For example, in Fig. \ref{fig5}.(c), the lights are fed in
evenly from all the waveguides, the initial state is $\left|\Psi_{g}\left(0\right)\right\rangle =\left[\begin{array}{cccc}
1 & 1 & \ldots & 1\end{array}\right]^{T}$. Our plots show the pumping is due to the edge states. In Fig.\ref{fig5}.(d-f),
the light is fed in from the most left unit cell. As initially in
the range $0\sim T_{1}$, there is no edge mode at the left edge (Fig.
\ref{fig3}), the light fed in from the most left are carried by the
bulk modes, which have completely decayed at the end of the time range
$0\sim T_{1}$. Nothing can be pumped or propagated in the two ranges
$T_{1}\sim T_{2}$ and $T_{2}\sim T_{3}$. In Fig. \ref{fig5}.(g),
though the light is fed in from the site of the most right unit cell,
it is still totally decayed in the range $0\sim T_{1}$. This is because
the right edge states have no distributions at the site $A$ (Fig.
\ref{fig3}.(e), (f)), and the light is still carried by the bulk
modes. The pumping process can be clearly seen from Fig.\ref{fig5}.(h-i):
after a few propagation length in $0\sim T_{1}$, the small portion
of bulk states are decayed while the edge modes are clearly left at
the sites $B,C$ (the two bright waveguides on the right). The pumping
from right edge to left edge can be seen in time range $T_{1}\sim T_{2}$.
After pumping, at the time range $T_{2}\sim T_{3}$, the light propagates
at the left sides. Here one thing should be mentioned that there is
no distribution of light on the waveguide $\mathcal{A}$ in the time
range $0\sim T_{1}$ and the distribution exists in the time range
$T_{2}\sim T_{3}$. This is because the waveguide $\mathcal{A}$ connects
site $A$ of initial unit cell and site $C$ of finite cell, and the
edge modes only exist at site $B,C$. This situation is reversed for
waveguide $\mathcal{C}$. Let's back to Fig. \ref{fig5}.(c), when
the lights are evenly fed in from all the waveguides, after long time
evolution, at the range $T_{2}\sim T_{3}$, the distribution is typical
like that of edge modes - although in contrast to Fig.\ref{fig5}.(h-i),
there is considerable loss output.

We'd like to strengthen that the imaginary parts of the on-site energies
$\varepsilon_{0,\mathcal{A}}$, $\varepsilon_{0,\mathcal{B}}$, $\varepsilon_{0,\mathcal{C}}$
are important to get the clear signal of pumping. In Fig. \ref{fig5}.(j-k),
the imaginary part of all the on-site energies are set zero, $\varepsilon_{0,\mathcal{A}}=\varepsilon_{0,\mathcal{B}}=\varepsilon_{0,\mathcal{C}}=0$
so that the pumping is adiabatic. This change has no remarkable effect
on the real part of the spectrum and the eigen states. The only change
is that the imaginary parts of all eigenvalues are exactly zero. However,
this change changes the pumping considerably. For Fig. \ref{fig5}.(j),
the light is fed in from the the most left wave guide $\mathcal{B}$,
the bulk states give a noisy signal after propagation, whereas in
Fig. \ref{fig5}.(e) we have no signal due to decay. For Fig. \ref{fig5}.(k),
the light is input from the most right waveguide $\mathcal{B}$, we
can seen relatively stronger signal at the right edge before pumping
and the relatively stronger signal at the left edge after pumping.
However, as compared to Fig. \ref{fig5}.(h), the signals are very
noisy due to contributions from bulk states.

\section{Enhanced pumping}

As discussed in above, the positive and negative imaginary parts of
the on-site energies completely change the nature of pumping by wiping
out the noisy contributions of bulk states. Further more, Fig.\ref{fig5}(c,h-i)
show that the light transmissions is strengthened during the pumping:
the intensities of the left edge modes after pumping are stronger
then initial edge modes at the right side.

The enhanced pumping can be well explained by the instantaneous eigen
spectrum of the Hamiltonian $H\left(t\right)$ for the system of finite
length. By solving $H\left(t\right)\varphi_{i}\left(t\right)=E_{i}\left(t\right)\varphi_{i}\left(t\right)$,
the spectrum $E_{i}\left(t\right)$ as a function of $\tilde{t}/T_{0}$
(with $T_{0}=T_{2}-T_{1}$ and $\tilde{t}=t-T_{1}$) is shown in Fig.
\ref{fig6}(a)-(b). Only the half period $\frac{\pi\tilde{t}}{T_{0}}\in\left[0,\pi\right]$
involved in pumping is shown, spectrum of another half period $\frac{\pi\tilde{t}}{T_{0}}\in\left[\pi,2\pi\right]$
can be easy get from the fact $H\left(\tilde{t}\right)=H\left(2\pi-\tilde{t}\right)$
(see (\ref{eq:6a}) and (\ref{eq:6b})). The two extra modes due to
the boundaries are marked by blue dashed line and red line. In the
real spectrum, they lie in the bottom of upper band and the top of
the bottom band; in the imaginary spectrum, the two bands are overlapped
and away from the bulk states. In the full time interval, the two
extra modes can not always be treated as edge modes. At $\tilde{t}=T_{0}/2$,
the real edge bands merge into the real bulk bands, the corresponding
eigenstates are also non-localized modes (Fig. \ref{fig6}(d)). This
merger leads to some transfer of population from edge modes to bulk
modes. The speciality of the small range around $\tilde{t}=T_{0}/2$
can also be seen from the imaginary part of eigen energies. At most
time range, the imaginary parts of the two extra bands are zero while
the imaginary parts of bulk modes are smaller then zero. However,
in the small range around $\tilde{t}=T_{0}/2$ and $\tilde{t}=3T_{0}/2$,
though the imaginary parts of the extra modes are still different
from the bulk modes, both the two extra modes and some bulk modes
gain the positive imaginary energies. The system thus gains energy
in this small range. As the imaginary parts of bulk states change
to negative after this small period, the energy gain by the bulk states
are decayed to the environment soon. However, as the imaginary part
of the two extra modes now becomes zero, the energy gained by the
extra modes is retained.

As total, a competing mechanism is introduced in the small range of
merging process around $\tilde{t}=T_{0}/2$: the extra modes may loss
energy to the bulk states, which is decayed in the following process;
it may also gain energy from the gain medium. Our numerical results
show that the signal of the output left edge mode can be enhanced
if the tuning range $T_{1}\sim T_{2}$ is made long enough. In such
a case, the extra modes can stay at the small range around $T_{0}/2$
long enough, so that the energy gaining from the gain medium can be
bigger then the energy losing to the bulk states. The situation can
be clearly seen for a wide sample. In Fig. \ref{fig6}(g)-(h), if
we choose $T_{0}=600/h_{0}$, the power gained is not strong enough
to fully send the edge mode from left to right, the strength of output
left edge modes are weaker then the input right edge modes; if we
double the tuning range so that $T_{0}=1200/h_{0}$, output left edge
modes are much brighter then the input right edge modes. Through this
way, we may enhance well localized states.

\section{Non-reciprocity}

\begin{figure}[t]
\begin{centering}
\includegraphics[width=1\columnwidth]{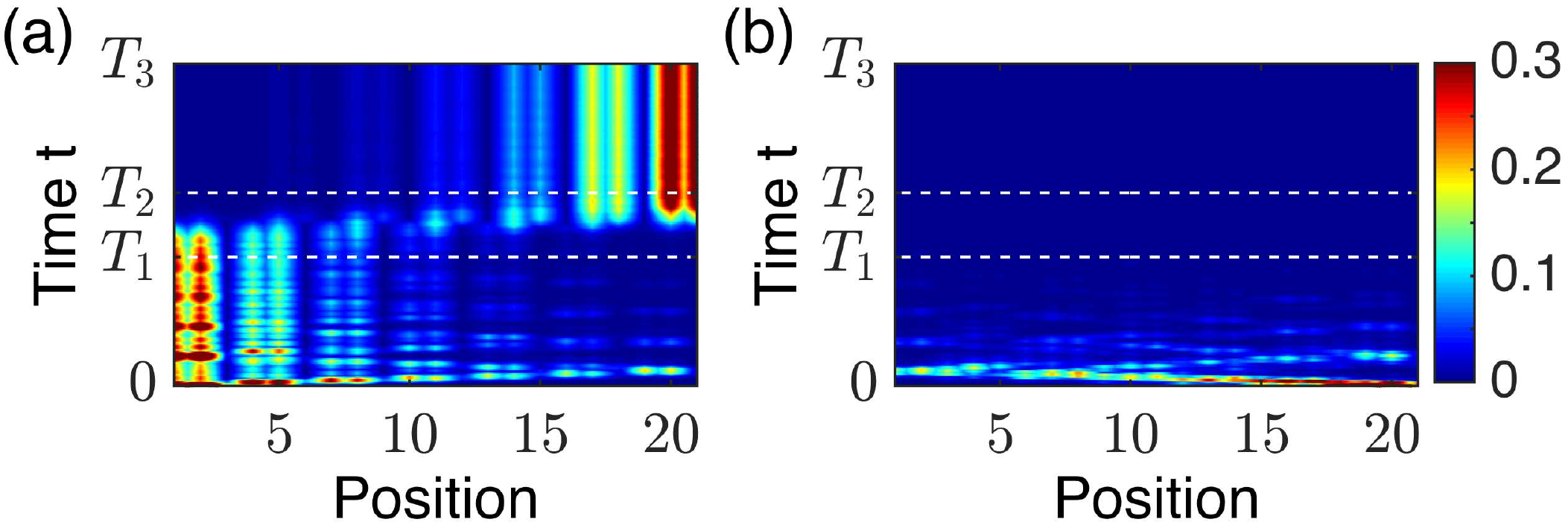}
\par\end{centering}
\caption{\label{fig7} Non-reciprocal property of light pumping. The waveguides
set (of Fig. \ref{fig5}.(a)) are up-side down so that $\alpha^{+}$
phase is at the bottom and the $\alpha^{-}$ phase is at the top.
Light is fed in from the $\alpha^{+}$ phase: (a), from the most left
wave guide $\mathcal{B}$; (b), from the most right waveguide $\mathcal{B}$.
The non-reciprocity is clearly seen by comparing Fig. \ref{fig7}a
(\ref{fig7}b) with Fig. \ref{fig5}e (\ref{fig5}h).}
\end{figure}

Our trimer lattice exhibits a very important property namely non-reciprocity
in propagation. More specifically, the pumping process between two
reversed phase $\alpha^{-}$ and $\alpha^{+}$ is non-reciprocal.
In Fig. \ref{fig7}, the waveguides set of Fig. \ref{fig5}.(a) are
made up-side down, and the light is fed in from the $\alpha^{+}$
phase. In Fig. \ref{fig7}.(a), the light is fed in from most left
$\mathcal{B}$ wave guide, we can witness the pumping of light from
the left to the right. However, as compared to Fig. \ref{fig5}.(e),
the light fed in the same waveguide from $\alpha^{-}$ phase can not
be pumped. The same situation happens when input light from the most
right waveguide: there is the pumping from the right to the left in
the Fig. \ref{fig5}.(h) when the light is input from $\alpha^{-}$
phase; while the light fast decays if the light is input from the
$\alpha^{+}$ phase (\ref{fig7}.(b)). The non-reciprocal property
is due to breaking of the vertical inversion symmetry $P_{y}$, which
is also equivalent to the broken time reversal symmetry. However,
the system still has the $\pi$ rotation symmetry or equivalently
the combination of the vertical and the horizontal inversion symmetries
$P_{x}P_{y}$, which makes Fig. \ref{fig7}.(a) equivalent to Fig.
\ref{fig5}.(h), and Fig. \ref{fig7}.(b) is equivalent to Fig. \ref{fig5}.(e)
after left-right reflection. It should be noted that we produce non-reciprocity
using a linear system (\ref{eq:1}), which is quite distinct from
several other recent approaches which use nonlinear optical methods
\cite{XL2014,zongfu2009,chunhuadong2014,JunHwan2015,Fan2012,Ganainy2013}.

\section{Discussion and OutLook}

We have presented a detailed study of the new phases which can arise
in trimer lattices. We specifically emphasize the new phases occurring
in finite systems. By studying 1D trimer lattices we reported edge
modes and new phases characterized by Berry phases which are piecewise
continuous rather than discrete numbers as in case of topological
phases. The phase transition occurs at the discontinuity point. We
discussed how trimer lattices can be used to have a 2D realization
with phases characterized by very specific 2D Berry phases of half
period. These characteristic Berry phases change smoothly within a
phase while change discontinuously at the transition point. We further
demonstrated the existence of adiabatic pumping for each phase and
gain assisted enhanced pumping. The non-reciprocity of the pumping
process makes the system a good optical diode. The results apply to
both electron and photon transport. As discussed in text it is easier
to realize the photon transport by using a system of waveguides and
we specifically took advantage of adding gain and loss in the waveguides.
Photonic lattices provide a new platform for the study of the phases
of matter.

In summary our work gives a new paradigm of phases and phase transitions.
Other phases may be found in similar ways. As an application, we present
a new mechanism for diode action which is based on symmetry. The current
research on trimer lattices can be combined with defects which are
expected to yield much richer physics.
\begin{acknowledgments}
One of us acknowledges his earlier association with Oklahoma State
University and the use of some of the facilities.
\end{acknowledgments}

\appendix*

\section{Design the experiment for pumping}

A 2D system can be realized by introducing time-dependent changes
in 1D trimer lattices. The time parameter basically gives the system
another dimension. This is especially realized in the context of photonic
lattices i.e. an array of waveguides as shown in Fig. \ref{fig5}.(a)\cite{Yaacov2012,H.Schomerus2013}.
By arranging the distance among waveguides and the coating media within
each waveguide, we may form the 1D unit cell that contains any number
of sites to model the 1D system. If we smoothly bend the waveguides
and slowly change the coating media within waveguides periodically,
the light propagates along the waveguides is described by a 2D Hamiltonian.
As the number of waveguide is always finite, non-trivial 2D topology
can be observed by the fact that light is pumped from one edge to
another during the propagation. This is the so-called adiabatic pumping.

One drawback of the physics of propagating light is that there is
no analog of the Fermi surface and filled bands. All the non-zero
component eigen modes of initial state $\Psi\left(0\right)$ contribute
to the propagating state $\Psi\left(t\right)$. Even when the initial
state $\Psi\left(0\right)$ is perfectly localized at one edge, bulk
states may still have non-zero probabilities at this edge. After propagation,
the contribution from bulk states can blur the contribution from edge
states. This problem is overcome by introducing small non-Hermitian
on-site energies \cite{H.Schomerus2013}. The key point is to tune
the parameter such that all the eigen-energies of bulk states have
negative imaginary parts, while the eigen-energies of edge modes are
exactly real. When the light propagates for long enough time, all
the bulk states decay and only the edge states are left. This results
in the clear signal of pumping light.

Non-trivial topology is not the necessary condition for pumping. For
our system, both the 2D phases can give the pumping process. This
is because the lattice of 1D $\alpha^{+}$ phase is the reverse lattice
of $\alpha^{-}$ phase (if the lattice formed by unit cell $ABC$
is at $\alpha^{-}$ phase, the lattice formed by $CBA$ is at $\alpha^{+}$
phase). While the two edge modes of 1D $\alpha^{-}$ phase that $\alpha=-\alpha_{0}<0$
are localized at the right edge (Fig. \ref{fig3}), the edge modes
of $\alpha^{+}$ phase that $\alpha=\alpha_{0}$ are localized at
the opposite edge. If we get the 2D system by tuning the parameters,
so that the lattice is slowly changed from $\alpha=-\alpha_{0}$ to
$\alpha=\alpha_{0}$, the two extra modes may slowly move from the
right edge to the left edge. Such a pumping process is not affected
by $\beta$, thus exists for both $\beta>1$ and $\beta<1$ 2D phases.
Only the half period $-\pi\leq k_{y}\leq0$ so that $-\alpha_{0}\leq\alpha\leq\alpha_{0}$
(with $\alpha_{0}>0$) of $\mathcal{H}\left(\beta,k_{y},k_{x}\right)$
(Eq. \ref{eq:4}) is involved in the pumping process from $\alpha^{-}$
phase to $\alpha^{+}$ phase. This is consistent with the fact that
both 2D phases are characterized by the half period Berry phase $\tilde{B}_{2}$.
In the following half period $0\leq k_{y}\leq\pi$, pumping process
is inverted and lattice is slowly changed back from $\alpha^{+}$
phase to $\alpha^{-}$ phase.

In the main context, we set the system to check the pumping process
for $\beta<1$ phase for half period. With the objective to wipe out
the bulk modes signal, the system is designed as follows. The waveguides
are supposed to be long enough and are separated in three parts $0\leftrightarrow L_{1}$,
$L_{1}\leftrightarrow L_{2}$ and $L_{2}\leftrightarrow L_{3}$ (Fig.
\ref{fig5}.(a)) . The first part $0\leftrightarrow L_{1}$ is the
decaying part, the distances and coating media of waveguides are fixed
so that waveguides set forms the trimer lattice of $\alpha^{-}$ phase.
Long length of this part guarantee that bulk states with negative
imaginary parts are fully decayed. In the part $L_{2}\leftrightarrow L_{3}$,
the parameters are set so that waveguides set forms reverse lattice
of the first part (see upper of Fig. \ref{fig5}.(a)), i.e. they are
at the $\alpha^{+}$ phase. The middle part $L_{1}\leftrightarrow L_{2}$
is the tuning part, the distances and coating media are slowly changed
from $\alpha^{-}$ phase to $\alpha^{+}$ phase. Long length of this
part guarantees a smooth tuning, which wipes out other non-adiabatic
factors that do not belong to the system. In general we need to bend
waveguides smoothly and slowly change the coating medium to achieve
tuning. Comparing the unit cell $ABC$ of $\alpha^{-}$ phase and
$CBA$ of $\alpha^{+}$ phase, the tuning part from lattice of $\alpha^{-}$
phase to the reverse lattice of $\alpha^{+}$ phase can be simplified
if we choose $\varepsilon_{0,A}=\varepsilon_{0,C}$. We then only
need to tune $h_{\mathcal{AB}}$ from $h_{AB}$ to $h_{BC}$ and $h_{\mathcal{BC}}$
from $h_{BC}$ to $h_{AB}$. Here the calligraphic symbol $\mathcal{A},\mathcal{B},\mathcal{C}$
are used to mark the different waveguides of one unit cell. Experimentally,
this can be done by bending waveguide $\mathcal{B}$ so that the distances
$l_{\mathcal{AB}}$ and $l_{\mathcal{BC}}$ are exchanged (Fig. \ref{fig5}.(a)).


\begin{thebibliography}{10}
\bibitem{Berry1980}M. V. Berry, \textsl{Exact Aharonov-Bohm wavefunction
obtained by applying Dirac's magnetic phase factor}, Eur. J. Phys.
\textbf{1}, 240 (1980).

\bibitem{Berry1984}M. V. Berry, \textsl{Quantal Phase Factors Accompanying
Adiabatic Changes}, Proc. R. Soc. Lond. A \textbf{392}, 45 (1984).

\bibitem{WilczekZee1984}Frank Wilczek and A. Zee, \textsl{Appearance
of Gauge Structure in Simple Dynamical Systems}, Phys. Rev. Lett.
\textbf{52}, 2111 (1984).

\bibitem{TKNN1982}D. J. Thouless, M. Kohmoto, M. P. Nightingale,
and M. den Nijs, \textsl{Quantized Hall Conductance in a Two-Dimensional
Periodic Potential}, Phys. Rev. Lett. \textbf{49}, 405 (1982).

\bibitem{Kohmoto1984} M. Kohmoto, \textsl{Topological invariant and
the quantization of the Hall conductance}, Annals of Physics \textbf{160},
343 (1985).

\bibitem{Class}Alexei Kitaev, \textsl{Periodic table for topological
insulators and superconductors}, AIP Conf. Proc. \textbf{1134}, 22
(2009).

\bibitem{Class1}Andreas P. Schnyder, Shinsei Ryu, Akira Furusaki,
and Andreas W. W. Ludwig, \textsl{Classification of topological insulators
and superconductors in three spatial dimensions}, Phys. Rev. B\textbf{
78}, 195125 (2008).

\bibitem{Class2}Shinsei Ryu, Andreas P Schnyder Akira Furusaki and
Andreas W W Ludwig, \textsl{Topological insulators and superconductors:
tenfold way and dimensional hierarchy}, New Journal of Physics \textbf{12},
065010 (2010).

\bibitem{Alexandra2014}A. Alexandradinata, Xi Dai, and B. Andrei
Bernevig, \textsl{Wilson-loop characterization of inversion-symmetric
topological insulators}, Phys. Rev. B \textbf{89}, 155114 (2014).

\bibitem{Alexandra2016}A. Alexandradinata, Zhijun Wang, and B. Andrei
Bernevig, \textsl{Topological Insulators from Group Cohomology}, Phys.
Rev. X \textbf{6}, 021008 (2016).

\bibitem{HasanReview2010}M. Z. Hasan and C. L. Kane, \textsl{Colloquium:
Topological insulators}, Rev. Mod. Phys. \textbf{82}, 3045 (2010).

\bibitem{QiReview2011}Xiao-Liang Qi and Shou-Cheng Zhang, \textsl{Topological
insulators and superconductors}, Rev. Mod. Phys. \textbf{83}, 1057
(2011).

\bibitem{shenbook2012}Shun-Qing Shen, \textit{Topological Insulators}
(Springer 2012).

\bibitem{bernevigbook2013}B. Andrei Bernevig with Taylor L. Hughes,
\textit{Topological Insulators and Topological Superconductors} (Princeton
University Press 2013).

\bibitem{XL2012}Xuele Liu, Qing-feng Sun, and X. C. Xie, \textsl{Topological
system with a twisting edge band: A position-dependent Hall resistance},
Phys. Rev. B \textbf{85}, 235459 (2012).

\bibitem{Hua2009}Hua Jiang, Lei Wang, Qing-feng Sun, and X. C. Xie,
\textsl{Numerical study of the topological Anderson insulator in HgTe/CdTe
quantum wells}, Phys. Rev. B \textbf{80}, 165316 (2009).

\bibitem{nagaosa2013}Motohiko Ezawa, Yukio Tanaka \& Naoto Nagaosa,
\textsl{Topological Phase Transition without Gap Closing}, Scientific
Reports \textbf{3}, 2790 (2013).

\bibitem{Rui2011}Rui Yu, Xiao Liang Qi, Andrei Bernevig, Zhong Fang,
and Xi Dai, \textsl{Equivalent expression of $\mathbb{Z}_{2}$ topological
invariant for band insulators using the non-Abelian Berry connection},
Phys. Rev. B \textbf{84}, 075119 (2011).

\bibitem{Vanderbilt2014}Maryam Taherinejad, Kevin F. Garrity, and
David Vanderbilt, \textsl{Wannier center sheets in topological insulators},
Phys. Rev. B \textbf{89}, 115102 (2014).

\bibitem{Vanderbilt2015}Maryam Taherinejad and David Vanderbilt,
\textsl{Adiabatic Pumping of Chern-Simons Axion Coupling}, Phys. Rev.
Lett. \textbf{114}, 096401 (2015).

\bibitem{Thouless1983}D. J. Thouless, \textsl{Quantization of particle
transport}, Phys. Rev. B \textbf{27}, 6083 (1983).

\bibitem{Qi2008}Xiao-Liang Qi, Taylor L. Hughes, and Shou-Cheng Zhang,
\textsl{Topological field theory of time-reversal invariant insulators},
Phys. Rev. B \textbf{78}, 195424 (2008).

\bibitem{Nicolo2012}Nicolò Spagnolo, Lorenzo Aparo, Chiara Vitelli,
Andrea Crespi, Roberta Ramponi, Roberto Osellame, Paolo Mataloni and
Fabio Sciarrino, \textsl{Quantum interferometry with three-dimensional
geometry}, Scientific Reports \textbf{2}, 862 (2012).

\bibitem{Chaboyer2015}Zachary Chaboyer, Thomas Meany, L. G. Helt,
Michael J. Withford \& M. J. Steel, \textsl{Tunable quantum interference
in a 3D integrated circuit}, Scientific Reports \textbf{5}, 9601 (2015).

\bibitem{Robert2015}Robert Keil, Changsuk Noh, Amit Rai, Simon St¸tzer,
Stefan Nolte, Dimitris G. Angelakis, and Alexander Szameit, \textsl{Optical
simulation of charge conservation violation and Majorana dynamics},
Optica \textbf{2}, 454 (2015).

\bibitem{XL2014}Xuele Liu, Subhasish Dutta Gupta and G. S. Agarwal,
\textsl{Regularization of the spectral singularity in $\mathcal{PT}$-symmetric
systems by all-order nonlinearities: Nonreciprocity and optical isolation},
Phys. Rev. A \textbf{89}, 013824 (2014).

\bibitem{zongfu2009}Zongfu Yu and Shanhui Fan, \textsl{Complete optical
isolation created by indirect interband photonic transitions}, Nature
Photonics \textbf{3}, 91 (2009).

\bibitem{chunhuadong2014}Chun-Hua Dong, Zhen Shen, Chang-Ling Zou,
Yan-Lei Zhang, Wei Fu \& Guang-Can Guo, \textsl{Brillouin-scattering-induced
transparency and non-reciprocal light storage}, Nature Communications
\textbf{6}, 6193 (2015).

\bibitem{JunHwan2015}JunHwan Kim, Mark C. Kuzyk, Kewen Han, Hailin
Wang and Gaurav Bahl, \textsl{Non-reciprocal Brillouin scattering
induced transparency}, Nature Physics \textbf{11}, 275 (2015).

\bibitem{Fan2012}L. Fan, J. Wang, L. T. Varghese, H. Shen, B. Niu,
Y. Xuan, A. M. Weiner, and M. Qi, \textsl{An All-Silicon Passive Optical
Diode}, Science \textbf{335}, 447 (2012).

\bibitem{Ganainy2013}R. El-Ganainy, A. Eisfeld, Miguel Levy, and
D. N. Christodoulides, \textsl{On-chip non-reciprocal optical devices
based on quantum inspired photonic lattices}, Appl. Phys. Lett. \textbf{103},
161105 (2013).

\bibitem{Yaacov2012}Yaacov E. Kraus, Yoav Lahini, Zohar Ringel, Mor
Verbin, and Oded Zilberberg, \textsl{Topological States and Adiabatic
Pumping in Quasicrystals}, Phys. Rev. Lett.\textbf{ 109}, 106402 (2012).

\bibitem{H.Schomerus2013}Henning Schomerus, \textsl{Topologically
protected midgap states in complex photonic lattices}, Optics Lett.
\textbf{38}, 1912 (2013).

\bibitem{Demange2012}Gilles Demange and Eva-Maria Graefe, \textsl{Signatures
of three coalescing eigenfunctions}, J. Phys. A: Math. Theor. \textbf{45},
025303 (2012).

\bibitem{AABook2003}A. P. Seyranian and A. A. Mailybaev, \textit{Multiparameter
Stability Theory with Mechanical Applications} (World Scientific Publishing
Co. Pte. Ltd., 2003).

\bibitem{Heeger1988}A. J. Heeger, S. Kivelson, J. R. Schrieffer,
and W. -P. Su, \textsl{Solitons in conducting polymers}, Rev. Mod.
Phys. \textbf{60}, 781 (1988).

\bibitem{Purcell1951}E. M. Purcell and R. V. Pound, \textsl{A Nuclear
Spin System at Negative Temperature}, Phys. Rev. \textbf{81}, 279
(1951).

\bibitem{Taylor2011}Taylor L. Hughes, Emil Prodan, and B. Andrei
Bernevig, \textsl{Inversion-symmetric topological insulators}, Phys.
Rev. B \textbf{83}, 245132 (2011).
\end{thebibliography}
\end{document}